\ifdefined\XeTeXrevision\else\pdfoutput=1\fi
\documentclass[11pt,a4paper]{article}

\usepackage[T1]{fontenc}
\usepackage{lmodern}
\usepackage[margin=2.5cm]{geometry}
\usepackage{amsmath,amssymb}
\usepackage{graphicx}
\usepackage{booktabs}
\usepackage{longtable}
\usepackage{array}
\usepackage{xcolor}
\usepackage{cite}
\usepackage{placeins}
\usepackage{pgfplots}
\pgfplotsset{compat=1.18}
\usepackage[colorlinks=true,linkcolor=blue!50!black,citecolor=blue!50!black,urlcolor=blue!50!black,pdftitle={Time-weighted energy correlators are infrared unsafe in gauge theories},pdfauthor={Siqing Yu}]{hyperref}

\numberwithin{equation}{section}
\newcommand{\vev}[1]{\langle #1\rangle}
\newcommand{\artanh}{\mathop{\mathrm{artanh}}}
\newcommand{\trQ}{\mathrm{tr}\,Q^2}
\newcommand{\calI}{\mathcal{I}}
\newcommand{\calM}{\mathcal{M}}
\newcommand{\calA}{\mathcal{A}}
\newcommand{\calC}{\mathcal{C}}
\newcommand{\calD}{\mathcal{D}}
\newcommand{\calO}{\mathcal{O}}
\newcommand{\Li}{\mathrm{Li}_2}

\newcommand{\dint}[1]{\int\!\frac{d#1}{2\pi i}}
\newcommand{\Nfour}{\mathcal{N}=4}

\definecolor{cA}{RGB}{0,0,0}
\definecolor{cB}{RGB}{0,114,178}
\definecolor{cC}{RGB}{0,140,120}
\definecolor{cD}{RGB}{213,94,0}
\pgfplotsset{every axis/.append style={
  width=0.47\textwidth, height=0.36\textwidth,
  tick label style={font=\scriptsize}, label style={font=\footnotesize},
  legend style={font=\scriptsize,draw=none,fill=none},
  legend cell align=left, line width=0.5pt, tick align=inside,
  every axis plot/.append style={line width=0.8pt,no markers}}}

\begin{document}

\hypersetup{pageanchor=false}
\begin{titlepage}
\begin{center}
\vspace*{1.5cm}
{\LARGE\bfseries Time-weighted energy correlators are infrared unsafe in gauge theories\par}
\vspace{1.2cm}
\renewcommand{\thefootnote}{\fnsymbol{footnote}}
{\large Siqing Yu\footnote{\href{mailto:yu@siqing.ai}{\texttt{yu@siqing.ai}}}\par}
\vspace{0.3cm}
{San Diego, California, USA\par}
\end{center}
\vspace{1.0cm}
\begin{abstract}
\noindent
Energy correlators are infrared safe because each detector weights a parton by its energy, but a weight in the arrival time makes them infrared unsafe in gauge theories. For the state created by a scalar operator that makes two partons at lowest order in a weakly coupled conformal gauge theory, we show that the correlator of two time-weighted energy detectors (the light-ray operators $\int\!du\,u\,T_{uu}$) at separated directions diverges at two loops like the logarithm of the time window. The logarithm does not come from a single soft gluon, but from a soft parton pair with comparable energies, one parton in each detector; its coefficient is a closed-form function of the angle for any matter content. In $\Nfour$ super Yang--Mills theory, we recover this coefficient from the two-gluon Regge term of the known four-point function and derive a simple relation to the logarithm in the charge correlator. In the planar theory the leading logarithms resum to a power of the window, set by the BFKL intercept.

\end{abstract}
\end{titlepage}
\hypersetup{pageanchor=true}
\setcounter{footnote}{0}

{\setlength{\parskip}{0.1em}\tableofcontents}
\newpage

\section{Introduction and summary}
\label{sec:intro}

A detector at future null infinity integrates a component of the stress tensor or of a current along a null generator. An energy detector weights each parton by its energy, and correlators of such detectors are the basic infrared- and collinear-safe observables: any safe observable can be approximated by them \cite{SveshnikovTkachov1996,Tkachov1997,Komiske2018}, and they are finite in perturbation theory \cite{HofmanMaldacena2008,BHKSZ2014a,BHKSZ2014b,BHKSZ2014c} and well defined in conformal field theory \cite{KKSZ2020a}. That argument concerns functions of the energies and directions of the final state, so it says nothing about a weight in the arrival time. The simplest such weight gives the first moment of the energy flux in the retarded time, $K(n)=\int du\,u\,E(u,n)$, which belongs to the light-ray algebras \cite{CordovaShao2018,HP2025,HP2026} and is the first frequency derivative of frequency-resolved detectors \cite{KSZ2022}.

The literature contains a qualitative warning about time-weighted detectors. Korchemsky and Zhiboedov caution that expanding frequency-resolved detectors in frequency brings powers of the working time under the integral, which weakens convergence and eventually makes matrix elements divergent \cite{KZ2021}. Kologlu, Kravchuk, Simmons-Duffin and Zhiboedov give the counting: products of such moments converge absolutely in the Regge limit for intercepts $J_0$ below a value that falls with the powers of time, so that $\vev{KK}$ is marginal at $J_0=1$. They note that $J_0=1$ is expected in a gauge theory, where their proof of commutativity stops, but leave the full analysis beyond their scope \cite{KKSZ2020a}. For the charge--charge correlator, which is known to exist if $J_0<1$, they expect divergences in gauge theories in both the weak and the strong coupling expansions \cite{KKSZ2020b}. Korchemsky, Sokatchev and Zhiboedov identify infrared unsafety as the collider avatar of such ill-defined products at zero frequency, and they state it for the charge--charge correlation \cite{KSZ2022}.

The danger can be stated in one sentence. A soft quantum of energy $E$ arrives over a time of order $1/E$, so the product of time and energy is scale invariant, and a time-weighted energy detector counts soft quanta the way a charge detector does, with equal weight per octave of energy. Charge correlators are known to be infrared unsafe \cite{Chicherin2020}, because a soft gluon can split into a pair that both detectors catch \cite{Monni2025}. The same splitting makes jet flavor infrared unsafe in widely used jet algorithms \cite{BSZ2006}, and it was expected to make charge correlators more singular at small angles \cite{HofmanMaldacena2008}. At one loop a single soft gluon gives no logarithm, since $K$ annihilates its eikonal factor; instead, a logarithm needs a pair, which first appears at two loops. On such a pair $K$ acts on the two partons with opposite signs, which makes the coefficient of the logarithm negative at separated directions.

We compute what this implies at weak coupling, with scattering amplitudes. For the state created by a scalar operator that makes two partons at lowest order in a weakly coupled conformal gauge theory, $\vev{K(n_1)K(n_2)}$ at separated directions contains the logarithm of the detector time window at two loops, so the correlator is infrared unsafe. The logarithm comes from a soft parton pair with comparable energies, one parton in each detector, and its coefficient is a closed-form function of the angle for any matter content. This logarithm and its coefficient do not rely on conformal symmetry, which we use only to exclude a further logarithm that would depend on the position of the source. In $\Nfour$ super Yang--Mills (SYM) theory the coefficient equals the two-gluon Regge term of the known four-point function, and takes the form
\begin{equation}\label{eq:intro}
\begin{aligned}
\vev{K_1K_2}_{\ln}&=-\frac{a^2}{32\pi^2z}\Big[(3-z)\calI(z)-1-\tfrac12\ln z\Big]\ln\Lambda\\
&=\frac{3-z}{4}\,\frac{\vev{Q_1Q_2}_{\ln}}{\trQ}+\frac{a^2}{32\pi^2z}\Big(1+\tfrac12\ln z\Big)\ln\Lambda .
\end{aligned}
\end{equation}
Here $a=g^2N/4\pi^2$, $z=\sin^2(\theta_{12}/2)$ with $\theta_{12}$ the angle between the two detectors, $\ln\Lambda$ is the logarithm of the window in units of the inverse source energy, $\calI(z)=\int_0^{\pi/2}d\phi\,\phi/(\sin^2\phi+z\cos^2\phi)$, and the subscript $\ln$ marks the term proportional to $\ln\Lambda$. The charge correlator $\vev{Q_1Q_2}$ is that of two detectors of an $R$-symmetry charge $Q$ under which the hard partons are neutral, and $\trQ$ is the sum of $Q^2$ over the three complex scalars, equal to that over the four Weyl fermions. The second line makes quantitative the statement that $K$ counts soft quanta the way a charge detector does: the two logarithms differ by an elementary factor and an elementary term. The four-point function contains every contribution at two loops, so the agreement with its Regge term confirms that in $\Nfour$~SYM the soft pair gives the whole logarithm.

The two-loop logarithm found here fits the convergence condition of \cite{KKSZ2020a}, which allows a logarithm only for a Regge intercept $J_0\geq1$. By an argument from spin counting, only two gluons reach $J_0=1$ at this order. Both are consistent with the absence of the logarithm in $\phi^4$ theory, where the soft pair is power suppressed.

As an operator the time-weighted detector looks harmless. At zero frequency, smeared on the sphere, it is a self-adjoint one-body operator on Fock space that annihilates the vacuum, and it acts on a soft pair by shifting the logarithm of the ratio of the two energies. Yet when acting on the state created by the source, it produces a state whose squared norm diverges logarithmically in the infrared, unless the smearing is uniform. Resummation in the planar theory does not remove the divergence. Beyond two loops, the logarithms of the window exponentiate: on Regge partial waves the time weight acts as the Legendre operator plus one, and the leading logarithms resum to a power of the window with the BFKL intercept. We obtain the coefficient of the squared logarithm at three loops in closed form.

In $\Nfour$~SYM, a related two-loop logarithm appears in a different observable with the same Regge function: for scalar detectors, Bercini and Kumar find it from the Regge limit in the $\mathbf{84}$ channel \cite{BK2026}. However, we have not found in the literature an explicit demonstration of the infrared unsafety of the time-weighted energy correlator at two loops, the soft-pair identification for a time weight, the coefficients, their dependence on the matter content, the closed forms, or the relation to the charge correlator. Likewise, we have not found the leading-logarithm tower of a time-weighted detector correlator or the closed form of its coefficient at order $a^3$.

The rest of the paper is organized as follows. Section~\ref{sec:setup} defines the detectors and the state. Section~\ref{sec:twoloop} computes the logarithm at two loops for general matter content and shows that in a conformal theory the other regions give no further logarithm in their sum. Section~\ref{sec:regge} identifies it in $\Nfour$~SYM with the two-gluon Regge term, which gives \eqref{eq:intro}. Section~\ref{sec:operator} treats the time-weighted detector as an operator on Fock space and shows what the logarithm means for a measurement on a normalizable state. Section~\ref{sec:beyond} uses this Regge origin to resum the leading logarithms in planar $\Nfour$~SYM and gives the three-loop coefficient. Section~\ref{sec:discussion} states the limits of the computation, lists open questions and relates the results to other detectors and to the light-ray algebras. Technical details are in the appendices, in particular the tail of the two-loop Mellin amplitude, the kernels of the detectors at small frequency and the derivation of the three-loop coefficient.

\section{Detectors and the state}
\label{sec:setup}

We use the mostly plus metric, a null vector $n=(1,\vec n)$, retarded time $u$ and radius $r$. At future null infinity, we define $E(u,n)=\lim_{r\to\infty}r^2T_{uu}$ and $J(u,n)=\lim_{r\to\infty}r^2j_u$, and from them the three light-ray operators,
\begin{equation}\label{eq:detectors}
E(n)=\int\!du\,E(u,n),\qquad K(n)=\int\!du\,u\,E(u,n),\qquad Q(n)=\int\!du\,J(u,n).
\end{equation}
Smeared on the sphere, $E[f]=\int d^2n\,f\,E$ and $K[f]=\int d^2n\,f\,K$, and $K[1]$ is the dilatation. In correlators such as $\vev{K_1K_2}$, the subscript labels the direction of the detector: $K_i=K(n_i)$, and likewise for $Q_i$ and $E_i$. Everywhere else, $E_i$ is the energy of parton $i$.

The frequency-resolved detector is $E_\omega(n)=\int du\,e^{i\omega u}E(u,n)$, so that $K=-i\partial_\omega E_\omega|_{\omega=0}$. A logarithmic divergence of a product is cut off by a nonzero frequency or by a smooth time window $\chi(u/T)$ of width $T$, with $\chi(0)=1$, in both detectors. The windowed detector $K_\chi$ has the weight $W(u)=u\chi(u/T)$ in place of $u$. Besides its part that conserves the number of partons, it has pair terms $\hat a\hat a$ and $\hat a^\dagger\hat a^\dagger$, which annihilate or create two partons in the detector's direction with total energy of order $1/T$. For a source of energy $M$ we write
\begin{equation}\label{eq:logLambda}
\ln\Lambda\equiv\ln(MT).
\end{equation}
The scales $1/T$ and $\omega$ agree up to a constant inside the logarithm, and we write $\vev{\cdots}_{\ln}$ for the term of a correlator proportional to $\ln\Lambda$. The coefficient of this logarithm does not depend on the shape of the window: the time-domain form in section~\ref{sec:softpair} shows this for gluons and minimally coupled scalars, and for every species it follows from the kernels at first order (appendix~\ref{app:kernels}).

Each time weight lowers the convergence bound on the Regge intercept by one. In the condition of \cite{KKSZ2020a} for absolute convergence in the Regge limit, with $J_0$ as the Regge intercept, $\vev{EE}$ needs $J_0<3$, $\vev{EK}$ needs $J_0<2$ and $\vev{KK}$ needs $J_0<1$.

\paragraph{Action on partons.} Take massless partons of any helicity, with energies $E_i$ and directions $n_i$. At zero frequency the number-conserving part of a detector acts on the momentum-space amplitude $F$ parton by parton:
\begin{equation}\label{eq:Konparton}
E(n)=\sum_iE_i\,\delta^2(n,n_i),\qquad K(n)=-i\sum_i\delta^2(n,n_i)\,(E_i\partial_{E_i}+1).
\end{equation}
The action of $K$ reads as the energy times the arrival time, with $-i\partial_E$ as the time on the amplitude and the $+1$ from the operator ordering in the measure $E\,dE$. It obeys $[E(n),K(n')]=i\delta^2(n,n')E(n)$. For a parton of momentum $P$ in the state and $P'$ in the bra, the number-conserving part of $E_\omega$ has the kernel $\sqrt{PP'}$ for a gluon, $\tfrac12(P+P')$ for a Weyl fermion and $\sqrt{PP'}+\xi(P'-P)^2/\sqrt{PP'}$ for a scalar, with $\xi$ the improvement coefficient; at $\xi=1/6$ these are the detector weights of \cite{KSZ2022}, divided by $\sqrt{PP'}$. These agree through first order in $P'-P$, so the action of $K$ does not depend on the species. On a loop amplitude with an infrared regulator, $K$ also sees the scale of the regulator; our computations apply $K$ to tree amplitudes only.

Acting on a soft eikonal factor $1/E$, the operator $E\partial_E+1$ gives zero. This is why $K$ has no logarithm at one loop. But the operator does not annihilate the amplitude of a soft pair, which depends on the ratio of the two energies. On an amplitude with a soft pole, $K$ also gives a finite term that \eqref{eq:Konparton} does not contain. In the basis used here, with a source at the origin in the ket and a momentum eigenstate in the bra, the expectation value of $K(n)$ has an endpoint term at zero energy when $E_iF$ tends to a nonzero limit as $E_i\to0$. An integration by parts in the energy splits the expectation value into two terms: the bulk term is the action of $K$ in \eqref{eq:Konparton}, and the boundary term at $E_i=0$ is the endpoint term. At order $a$ the endpoint term equals $-i/2$ times the density of soft quanta in the direction $n$, per unit $\ln E$ and per unit solid angle, and together with the bulk term it reproduces the exact form of $\vev{E_1K_2}$ in section~\ref{sec:hardsoft}.

\paragraph{The state.} The state is created by a scalar operator $O$ of dimension $\Delta$ with momentum $q=(M,\vec0)$, normalized by its two-point function, and at lowest order it contains a color-singlet back-to-back pair of partons in a representation $R$ with momenta $p_{a,b}=\tfrac M2(1,\pm\hat n)$. We set $M=1$ where no confusion arises. Angular brackets $\vev{\cdot}$ on a function of the hard axis denote its average over $\hat n$ at fixed detector directions. The angle between the two detectors enters through $n_{12}\equiv-n_1\cdot n_2=2z$, with $z=\sin^2(\theta_{12}/2)$. In $\Nfour$~SYM, $R$ is the adjoint and the source is a $\mathbf{20'}$ operator, with $a=g^2N/4\pi^2$. There $C_R=C_A=N$, so the prefactor $g^4C_RC_A/16\pi^4$ of the two-loop formulas is $a^2$ for any $N$, not only in the planar limit. For general $R$ we write $a=g^2C_R/4\pi^2$ in one-gluon formulas.

At leading power in the soft energies the amplitude factorizes onto two Wilson lines along $n_a=(1,\hat n)$ and $n_b=(1,-\hat n)$, and the soft factor does not depend on the hard energies. In a stereographic coordinate $\zeta$ on the sphere, with $\zeta_a$ and $\zeta_b$ the images of $n_a$ and $n_b$, the probability of emitting one soft gluon is
\begin{equation}\label{eq:softrate}
dP=a\,\frac{dE}{E}\,\frac{d^2\zeta}{\pi}\,|S(\zeta)|^2,\qquad S(\zeta)=\frac1{\zeta-\zeta_a}-\frac1{\zeta-\zeta_b},
\end{equation}
summed over helicities.

\section{The logarithm at two loops}
\label{sec:twoloop}

Through order $a^2$, we show that in a conformal theory the only logarithm of the window in $\vev{K_1K_2}$ at separated directions comes from a soft pair of partons with comparable energies, one in each detector. The logarithm comes from pairs with energies between $1/T$ and $M$, where only the first order in the frequency matters, so the differences between species in the exact frequency kernels at higher order do not enter.

\subsection{The soft pair}
\label{sec:softpair}

The amplitude for a soft pair of gluons with momenta $k_1=E_1n_1$ and $k_2=E_2n_2$, emitted from the two Wilson lines, has two color structures. The part symmetric in the colors $c_1,c_2$ is independent emission, a product of two one-gluon eikonal currents with its color factor symmetrized. In the part with $f^{c_1c_2d}T^d$, where $T^d$ is the color charge of parton $a$ and $-T^d$ that of $b$, the pair couples in color to the Wilson lines as a single gluon $d$; we call this color structure connected. At leading soft power, a pair of matter partons comes from one soft gluon $d$, so it has only the connected structure, with the generator of its representation in place of $f^{c_1c_2d}$.

At leading soft power the amplitude of the pair is homogeneous of degree $-2$ in the two energies, like a product of two eikonal factors, so we separate their scale from their ratio, with $\lambda=\sqrt{E_1E_2}$ and $y=\ln(E_1/E_2)$. The double-soft currents \cite{CG2000a,CaronHuot2018} depend on the ratio only through the light-cone momentum fractions carried by parton 1, one along each Wilson line,
\begin{equation}
p=\frac{n_a\cdot k_1}{n_a\cdot(k_1+k_2)}=\frac{E_1A_1}{E_1A_1+E_2A_2},\qquad q=\frac{n_b\cdot k_1}{n_b\cdot(k_1+k_2)}=\frac{E_1B_1}{E_1B_1+E_2B_2},
\end{equation}
with $A_i=-n_a\cdot n_i$ and $B_i=-n_b\cdot n_i$, so that parton 2 carries $1-p$ and $1-q$, and $\partial_yp=p(1-p)$, $\partial_yq=q(1-q)$. In the connected color structure the amplitude is $F=s(y)/\lambda^2$, and in these variables the currents give
\begin{equation}\label{eq:softforms}
\begin{aligned}
s_{\rm sc}&=\frac{p-q}{n_{12}},\qquad s_{\rm f}=\frac{u_a\sqrt{p(1-p)}-u_b\sqrt{q(1-q)}}{n_{12}},\\
s_{\rm g}&=\frac{u_a^2(1-2p)-u_b^2(1-2q)}{2n_{12}}+\text{(independent of $y$)},
\end{aligned}
\end{equation}
for scalars, Weyl fermions and $g^+g^-$, respectively, where $u_{a,b}$ is the phase of $\langle1|n_{a,b}|2]$. Only the relative phase enters, $u_au_b^*=e^{i\phi}$, with $\phi$ the azimuth of $n_2$ from $n_1$ around the hard axis. The pair $g^+g^+$ and independent emission do not depend on $y$. The expressions for $s_{\rm sc}$ and $s_{\rm f}$ also follow directly from the eikonal current $j^\mu=n_a^\mu/(n_a\cdot k)-n_b^\mu/(n_b\cdot k)$, with $k=k_1+k_2$, through $j\cdot k_1=p-q$ and the identity $2n_{12}=A_1B_2+A_2B_1-2\cos\phi\sqrt{A_1A_2B_1B_2}$, while $s_{\rm g}$ follows from the full double-soft gluon current.

Since $F$ has degree $-2$ in the energies, $(E_1\partial_1+1)F=\partial_yF$ and $(E_2\partial_2+1)F=-\partial_yF$. Hence \eqref{eq:Konparton} acts on the pair as
\begin{equation}\label{eq:KKpair}
K_1K_2F=\partial_y^2F.
\end{equation}
With the phase space $E_idE_i\,d^2n_i$ per parton, and with $n_1$ and $n_2$ fixed by the two detectors, the contribution of the pair to $\vev{K_1K_2}$ is, up to a positive factor, $\int E_1dE_1\,E_2dE_2\,F^*K_1K_2F$, summed over colors, helicities and species. In the variables $y$ and $\lambda$ the measure is $E_1dE_1\,E_2dE_2=\lambda^3d\lambda\,dy$, and by \eqref{eq:KKpair} $F^*K_1K_2F=s^*\partial_y^2s/\lambda^4$, so the integral splits into $\int d\lambda/\lambda$ times $\int dy\,s^*\partial_y^2s$. An integration by parts in $y$ turns the second factor into $-\int dy\,|\partial_ys|^2$. Its boundary term $s^*\partial_ys$ vanishes as $y\to\pm\infty$, where one parton of the pair is much softer than the other, because $\partial_ys\to0$ there. Independent emission and the pair $g^+g^+$, which do not depend on $y$, therefore drop out. The scale integral $\int d\lambda/\lambda$ diverges logarithmically; cutting it off at $\lambda\sim1/T$ with the window and at $\lambda\sim M$ with the soft approximation then gives $\ln\Lambda$.

With the energy weights of $\vev{E(n_1)E(n_2)}$ in place of $K_1K_2$, the integrand is instead $E_1E_2|F|^2=|s|^2/\lambda^2$, which, with the measure, is $\lambda\,|s|^2\,d\lambda\,dy$ and converges as $\lambda\to0$; the logarithm therefore comes from the time weight. The sign can also be read off without integrating. Since $(E_1\partial_1+1)F=-(E_2\partial_2+1)F$, the sum $K_1+K_2$ annihilates the pair, so $K_2F=-K_1F$, and the integral is $-\int E_1dE_1\,E_2dE_2\,|K_1F|^2$, the same sum of squares. In other words, the two time-weighted energies are exactly anticorrelated in every configuration of the pair.

\paragraph{The time domain.} The window enters in the simplest way for gluons and minimally coupled scalars, where $E(u,n)$ on one parton is $1/(2\pi)$ times the square of a Fourier transform in the energy (appendix~\ref{app:kernels}). At leading soft power, up to the overall factor of \eqref{eq:KKsum}, the connected pair contributes $|\mathcal T|^2/(2\pi)^2$ to $\vev{E(u_1,n_1)E(u_2,n_2)}$, with the amplitude
\begin{equation}\label{eq:timeamp}
\mathcal T(u_1,u_2)=\int_0^\infty\!\!\int_0^\infty\!dE_1dE_2\,s(y)\,e^{-i(E_1u_1+E_2u_2)}=-\int_0^1\!\frac{dx\;s(y)}{(xu_1+(1-x)u_2-i0^+)^2},
\end{equation}
where $x=E_1/(E_1+E_2)$. Since $\mathcal T(Lu_1,Lu_2)=L^{-2}\,\mathcal T(u_1,u_2)$, in polar coordinates $(u_1,u_2)=L(\cos\gamma,\sin\gamma)$, where $d^2u=L\,dL\,d\gamma$, the integral of $u_1u_2|\mathcal T|^2$ over the times is $\int dL/L$ times
\begin{equation}\label{eq:timecoef}
\int_0^{2\pi}\!d\gamma\,\cos\gamma\sin\gamma\,|\mathcal T|^2_{L=1}=-(2\pi)^2\int\!dy\,|\partial_ys|^2 .
\end{equation}
It follows from Parseval's theorem, which turns the integral over the times into one over the energies. The factor $u_1u_2$ on $\mathcal T$ is $-\partial_1\partial_2$ on $s$, and $-\partial_1\partial_2s=\partial_y^2s/(E_1E_2)$, which is \eqref{eq:KKpair} again. If $s\to0$ at $y\to\pm\infty$, as $s_{\rm sc}$ does, no boundary terms arise at $E_1=0$ or $E_2=0$, and the integral over the times is
\begin{equation}\label{eq:Tnorm}
\int\!d^2u\,u_1u_2|\mathcal T|^2=(2\pi)^2\!\int\!\frac{dE_1dE_2}{E_1E_2}\,s^*\partial_y^2s=-(2\pi)^2\!\int\!d\ln\lambda\int\!dy\,|\partial_ys|^2 .
\end{equation}
Both sides diverge logarithmically: the left side is $\int dL/L$ times the left side of \eqref{eq:timecoef}, with $L$ in the role of $1/\lambda$. An envelope of large width in $\ln\lambda$ multiplying $s$ makes both finite. It removes the corner $E_1=E_2=0$, where the transform is defined only up to a constant, and it adds only $\pi^2\int dy\,|s|^2$ times the integral of its squared slope over $\ln\lambda$, which does not grow with the width. The sign comes from times of opposite signs: for $s=p-q$, the two quadrants with $\cos\gamma\sin\gamma>0$ give a positive part of the left side of \eqref{eq:timecoef}, and the two where $u_1$ and $u_2$ have opposite signs give a negative part that outweighs it, since the right side of \eqref{eq:timecoef} is negative.

A window with $\chi(0)=1$ ends the integral at $L\sim T$, and the soft approximation, valid for energies below $M$, ends it at $L\sim1/M$, which gives $\ln(MT)$ with a coefficient that does not depend on the window. The logarithm thus comes from the region of large times at fixed ratio, the Regge limit of the four-point function integrated along the detectors \cite{KKSZ2020a,KSZ2022,BK2026}. For fermions and for scalars with an improvement term the kernels differ at higher order in $1/(\lambda T)$, which changes only the constant.

For gluons $s$ tends to constants $s_\pm$ as $y\to\pm\infty$, and the integral over $\gamma$ in \eqref{eq:timecoef} then diverges at the axes $u_1=0$ and $u_2=0$, so it needs a cut symmetric about each axis. The part $s_-+(s_+-s_-)E_1/(E_1+E_2)$ of $s$ gives $\mathcal T$ the tails $-s_-/(u_1u_2)$ as $u_2\to0$ and $-s_+/(u_1u_2)$ as $u_1\to0$, and the transform of the rest grows only logarithmically at the axes. Near the axis $u_2=0$ the integrand is then $|s_-|^2/(\cos\gamma\sin\gamma)$, and near $u_1=0$ it is $|s_+|^2/(\cos\gamma\sin\gamma)$, up to integrable terms; each is odd across its axis. So each quadrant diverges logarithmically at both of its axes, with opposite signs in adjacent quadrants: with a cut symmetric about each axis the divergences cancel, and the four quadrants together converge to the right side of \eqref{eq:timecoef}. Damping $s$ at large $|y|$ by a factor $e^{-\epsilon|y|}$ is such a cut as $\epsilon\to0$: by Parseval's theorem it changes the right side only by terms of order $\epsilon$. Windows in the detectors, whether even in the time or centered away from $u=0$, give the same coefficient: each tail $-s_\mp/(u_1u_2)$ is a product of two eikonal factors, one for each detector, and gives no logarithm, since $K$ annihilates an eikonal factor.

\subsection{The logarithm for general matter content}
\label{sec:general}

\begin{figure}[t]
\centering
\begin{tikzpicture}
\begin{axis}[xlabel={$z$},ylabel={$\vev{G}$},xmin=0,xmax=1,ymin=0,ymax=1.8,legend style={at={(0.5,1.02)},anchor=south,legend columns=3}]
\addplot[cA] table[x=z,y=N4] {figures/fig1a.csv};
\addlegendentry{$\Nfour$}
\addplot[cD,dashed] table[x=z,y=BZ] {figures/fig1a.csv};
\addlegendentry{$(0,11)$}
\addplot[cB,densely dotted] table[x=z,y=gf] {figures/fig1a.csv};
\addlegendentry{$8\vev{g_f}$}
\addplot[cC,dashdotted] table[x=z,y=gg] {figures/fig1a.csv};
\addlegendentry{$2\vev{g_g}$}
\addplot[cA!55,dotted,line width=1pt] table[x=z,y=gs] {figures/fig1a.csv};
\addlegendentry{$6\vev{g_s}$}
\end{axis}
\end{tikzpicture}
\caption{The angular function $\vev{G}$ of \eqref{eq:KKgeneral} for $\Nfour$~SYM, with the shares of the species, and at the Banks--Zaks endpoint $(n_s,n_f)=(0,11)$ of the conformal line.}
\label{fig:KK}
\end{figure}
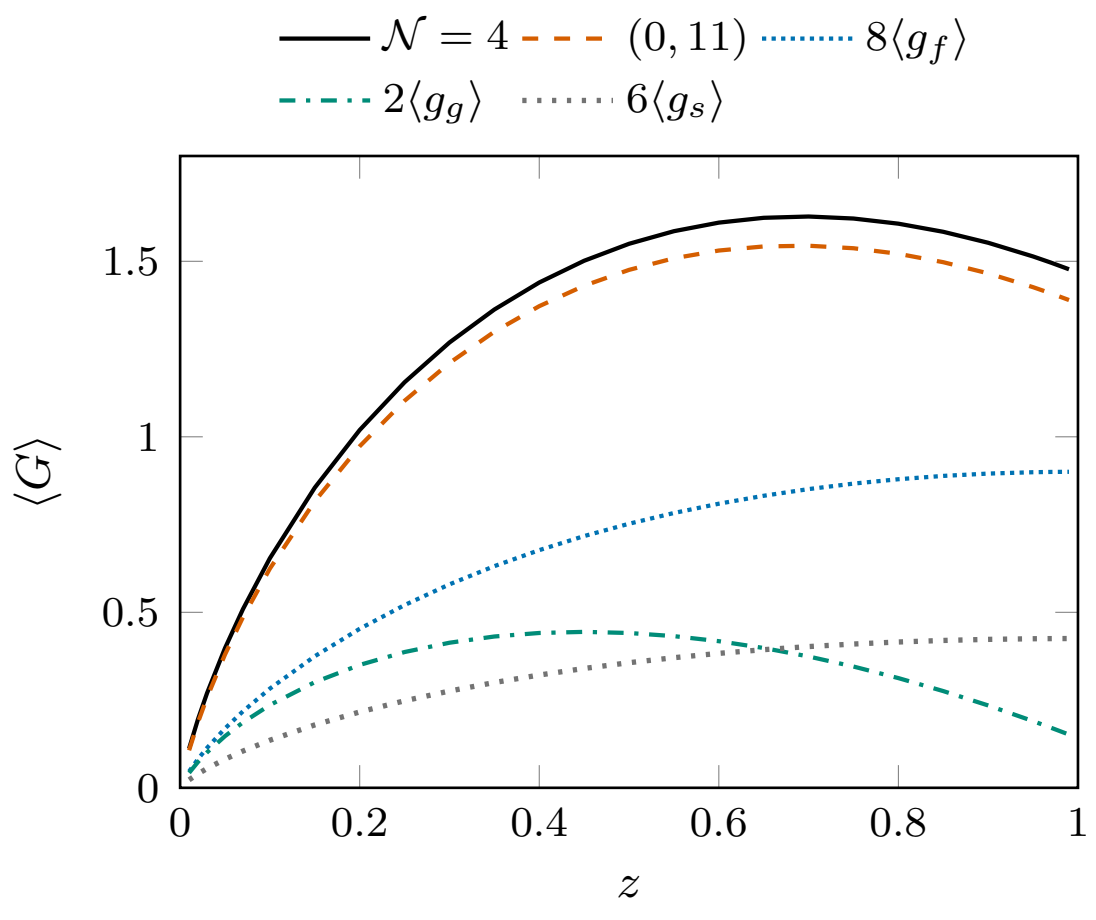

Let the matter content be measured by
\begin{equation}
n_s=\sum_{\text{real scalars}}\frac{T_r}{C_A},\qquad n_f=2\sum_{\text{Weyl}}\frac{T_r}{C_A},
\end{equation}
with $T_r$ the index of each representation, so that $\Nfour$~SYM has $(n_s,n_f)=(6,8)$. With them,
\begin{equation}\label{eq:KKsum}
\vev{K_1K_2}_{\ln}=-\frac{g^4C_RC_A}{16\pi^4}\,\frac1{16\pi^2}\,\Big\langle\int\!dy\,\big(n_s|\partial_ys_{\rm sc}|^2+n_f|\partial_ys_{\rm f}|^2+2|\partial_ys_{\rm g}|^2\big)\Big\rangle\ln\Lambda .
\end{equation}
The integrals over $y$ are elementary. The fractions $p$ and $q$ have the same form as functions of $y$ and differ only in their offsets, $y_a=\ln(A_1/A_2)$ and $y_b=\ln(B_1/B_2)$: $p=1/(1+e^{-y-y_a})$ and $q=1/(1+e^{-y-y_b})$. Since a shift of $y$ leaves the integrals unchanged, at fixed detector directions the integrals depend on the hard axis only through $\phi$ and $y_a-y_b=2b$. With
\begin{equation}
b=\frac12\ln\frac{A_1B_2}{A_2B_1},\qquad h_1(b)=\frac{b\coth b-1}{\sinh^2b},\qquad h_2(b)=-\partial_b^2\frac{2b}{\sinh b},
\end{equation}
the logarithm is
\begin{equation}\label{eq:KKgeneral}
\vev{K_1K_2}_{\ln}=-\frac{g^4C_RC_A}{16\pi^4}\,\frac{\vev{G}}{32\pi^2z^2}\,\ln\Lambda,\qquad G=n_sg_s+n_fg_f+2g_g,
\end{equation}
with one function for each species,
\begin{equation}\label{eq:species}
g_s=\frac{1-3h_1}{6},\qquad g_f=\frac{\tfrac23-\cos\phi\,h_2}{8},\qquad g_g=\frac{\tfrac13-\cos2\phi\,h_1}{2}.
\end{equation}
Recall that $z=\sin^2(\theta_{12}/2)$, with $\theta_{12}$ the angle between the two detectors. Since each of $g_s$, $g_f$ and $g_g$ is the integral of a square, $G\geq0$, with $G>0$ wherever $b\neq0$, so $\vev{K_1K_2}_{\ln}<0$ at every angle for every matter content. The average over the hard axis, derived in appendix~\ref{app:reduction}, is a closed form in the function $\calI(z)$:
\begin{equation}\label{eq:Gclosed}
\vev{G}=-\frac{c_{\calI}}{128z}\,\calI(z)+\frac{c_{\rm r}}{384z}-\frac{c_{\ln}}{256z}\,\ln z ,
\end{equation}
\begin{equation}
\begin{aligned}
c_{\calI}&=n_s(1+z)(3z^2-14z+3)+n_f(5z^3-21z^2+3z-3)+2(35z^3-75z^2+21z+3),\\
c_{\rm r}&=n_s(9-32z-9z^2)+n_f(8z-9-15z^2)+2(9+64z-105z^2),\\
c_{\ln}&=n_s(z-3)(3z-1)+n_f(1+z)(5z-3)+2(1+5z)(3+7z),
\end{aligned}
\end{equation}
where, with the Legendre chi function $\chi_2(x)=\tfrac12[\Li(x)-\Li(-x)]$,
\begin{equation}\label{eq:calI}
\calI(z)=\int_0^{\pi/2}\!\frac{\phi\,d\phi}{\sin^2\phi+z\cos^2\phi}=\frac{\chi_2(\sqrt z)-\ln\sqrt z\,\artanh\sqrt z}{\sqrt z}.
\end{equation}
At $(n_s,n_f)=(6,8)$ all the $1/z$ terms cancel, which leaves $\vev{G}=z[(3-z)\calI(z)-1-\tfrac12\ln z]$, the form that section~\ref{sec:regge} derives from the Regge term. This function is shown in figure~\ref{fig:KK}, with the shares of the species.

In $\Nfour$~SYM the soft-pair result also has a compact form as a single integral. With the energy asymmetry of the pair, $w=(E_1-E_2)/(E_1+E_2)=\tanh(y/2)$, and its velocity, $v=|\vec k_1+\vec k_2|/(E_1+E_2)$, which obeys $v^2=1-z(1-w^2)$, the average over the hard axis at fixed energy ratio gives
\begin{equation}\label{eq:KKw}
\vev{K_1K_2}_{\ln}=-\frac{a^2}{32\pi^2z}\int_{-1}^1\!dw\,\Big[(1+v^2)\frac{\artanh v}{v}-1\Big]\ln\Lambda .
\end{equation}
The function $\calI$ has an integral form of the same kind, $\calI(z)=\int_0^1dw\,\artanh(v)/v$. Writing $\artanh(v)/v=\int_0^1dr/(1-v^2r^2)$ and $\sin^2\phi=(1-r^2)/(1-r^2w^2)$ turns it into the integral over $\phi$ in \eqref{eq:calI}, and the closed form follows from $z\partial_z[\sqrt z\,\calI(z)]=-\sqrt z\ln z/[4(1-z)]$.

Two properties of $G$ follow from its structure. First, it is invariant under conformal maps of the sphere, because $b$ is the logarithm of the modulus of the cross ratio of the four directions $n_1,n_2,n_a,n_b$, and $\phi$ is its phase. Second, it is finite for any matter content: the integral over $y$ converges for each species. In the sum of squares the terms in $[p(1-p)]^2$ and $[q(1-q)]^2$ have the coefficient $(n_s-n_f+2)/n_{12}^2$, which vanishes in $\Nfour$~SYM. The finiteness does not rest on this cancellation: these terms are integrable, and since $\int dy\,[p(1-p)]^2=1/6$ they only add to $G$ the constant $(n_s-n_f+2)/6$.

\paragraph{The conformal line.} The one-loop beta function vanishes on the line $22=2n_f+n_s$, where $G=G_{\Nfour}+(n_f-8)(g_f-2g_s)$. An $\mathcal N=1$ theory has $n_f=2(1+x)$ and $n_s=2x$ with $x=\sum T_r/C_A$ over the chiral multiplets, and a vanishing one-loop beta function forces $x=3$. So every theory with $\mathcal N\geq1$ supersymmetry and vanishing one-loop beta function has $(n_s,n_f)=(6,8)$, and the angular function is common to all of them, with only the overall Casimir $C_R$ to distinguish them. At the free Banks--Zaks endpoint $N_f=11N_c/2$ \cite{BanksZaks1982}, with $(n_s,n_f)=(0,11)$, the function differs from that of $\Nfour$~SYM by at most 6\% for $0.01\le z\le0.99$.

\paragraph{Charge flux.} The same soft pair gives the known infrared unsafety of charge flux, with a closed form. Let $\calC_s$ and $\calC_f$ be the sums of $Q^2T_r/C_A$ over complex scalars and over Weyl fermions, and let the hard partons be neutral under $Q$. Then
\begin{equation}\label{eq:QQgeneral}
\vev{Q_1Q_2}_{\ln}=-\frac{g^4C_RC_A}{16\pi^4}\,\frac{1}{16\pi^2n_{12}^2}\,\big\langle2\calC_sU_s+2\calC_fU_f\big\rangle\ln\Lambda,
\end{equation}
\begin{equation}\label{eq:Us}
\begin{aligned}
U_s&=\int\!dy\,(p-q)^2=2b\coth b-2,\\
U_f&=\int\!dy\,\big|u_a\sqrt{p(1-p)}-u_b\sqrt{q(1-q)}\big|^2=2-\frac{2b\cos\phi}{\sinh b}.
\end{aligned}
\end{equation}
The averages over the hard axis are
\begin{equation}\label{eq:Uaverages}
\vev{U_s}=(1+z)\calI(z)-1+\tfrac12\ln z,\qquad\vev{U_f}=(3z-1)\calI(z)+1-\tfrac12\ln z.
\end{equation}
When $\calC_s=\calC_f$, the elementary terms cancel and $\vev{U_s+U_f}=4z\calI(z)$. In $\Nfour$~SYM, $\calC_s=\calC_f=\trQ$, and the charge correlator has the logarithm
\begin{equation}\label{eq:QQN4}
\vev{Q_1Q_2}_{\ln}=-\frac{a^2\,\trQ}{8\pi^2z}\,\calI(z)\ln\Lambda .
\end{equation}

\subsection{The other regions}
\label{sec:hardsoft}

To establish that the soft pair gives the full coefficient of $\ln\Lambda$, we account for the remaining regions. Table~\ref{tab:regions} summarizes the contributions of all regions through order $a^2$. Each detector acts on a hard parton, a soft parton or through its pair term; this covers six combinations, with the crosstalk of \cite{BHKSZ2014b,KSZ2022}, the exchange of a scalar at zero energy between the two detectors, as a separate effect. The first four rows are evaluated on explicit amplitudes; the other rows follow from general arguments. Those involving a detector on a hard parton, and the pair term with the other detector on a soft parton, take a longer argument, and we treat them here, together with the crosstalk.

\begin{table}[htbp]
\centering
\footnotesize
\begin{tabular}{@{}>{\raggedright}p{0.42\textwidth} >{\raggedright\arraybackslash}p{0.52\textwidth}@{}}
\toprule
Region & Logarithm \tabularnewline
\midrule
Soft pair, one parton in each detector, $E_1\sim E_2$ & all of the real part: \eqref{eq:KKgeneral} \tabularnewline
One soft gluon, $O(a)$ & none: $(E\partial_E+1)(1/E)=0$ \tabularnewline
Independent emission; $g^\pm g^\pm$ & none: no dependence on $E_1/E_2$ \tabularnewline
Strongly ordered soft pair & none: $K$ annihilates the softer factor \tabularnewline
Hard--soft & imaginary only: $-i\Delta\vev{E_1K_2}_{\rm hs}$ \tabularnewline
Pair term, other detector on a hard parton & imaginary only: $-i\Delta/2$ times a real share of $\vev{K}$ \tabularnewline
Hard--hard, soft radiation undetected & none: the frequencies reach only hard partons \tabularnewline
Pair term, other detector on a soft parton & none: a total derivative in the detected energy, finite limit \tabularnewline
Pair terms of both detectors & none: all energies $\lesssim1/T$, no dependence on $M$ \tabularnewline
Crosstalk: zero-energy scalar exchange & none: detected energies $\lesssim1/T$, other integrals converge \tabularnewline
\bottomrule
\end{tabular}
\caption{Regions that could give a logarithm of the window in $\vev{K_1K_2}$ at separated directions through order $a^2$. Pair terms are a windowed detector's $\hat a\hat a$ and $\hat a^\dagger\hat a^\dagger$ parts, with energies $\lesssim1/T$. In a conformal theory the sum of the two imaginary entries has no logarithm.}
\label{tab:regions}
\end{table}

\paragraph{Hard--soft region.} Here $K_1$ acts on a hard parton and $K_2$ on a soft one. The argument below has two conclusions: in a conformal theory the region and the pair term of $K_2$ give no logarithm in their sum, and for any matter content the region gives none in the real part of $\vev{K_1K_2}$. At order $a^2$ the region contains configurations with three hard partons, in which the soft gluon is at tree level and $K$ annihilates its eikonal factor, so they give no logarithm. With two hard partons, at leading soft power, $E$ on a hard parton is the constant $M/2=\tfrac12$ in units with $M=1$, and $K$ on the same parton is the constant $-i\Delta/2$, since the tree-level amplitude of the pair has degree $(\Delta-2)/2$ in each hard energy. Both constants then multiply the same quantity, the expectation of $K(n_2)$ in the Wilson-line state along $n_1$ and $(1,-\vec n_1)$, at fixed axis and with no average, so the ratio of the two correlators is that of the constants:
\begin{equation}\label{eq:hsidentity}
\vev{K_1K_2}_{\rm hs}=-i\Delta\,\vev{E_1K_2}_{\rm hs}.
\end{equation}
In $\Nfour$~SYM, with the energy correlator $\mathrm{EEC}(z)=8\pi^2\vev{E_1E_2}$, the correlator $\vev{E_1K_2}$ has an exact form, derived in appendix~\ref{app:identity} from the kernels of \cite{KSZ2022} at small frequency:
\begin{equation}\label{eq:EKidentity}
\vev{E_1K_2}=-\frac{i}{4\pi^2}\,\frac{1}{z^2(1-z)}\Big(1-\frac12X\partial_X\Big)\big[z^2(1-z)\,\mathrm{EEC}(z)\big],\qquad X=\frac{1-z}{z},
\end{equation}
which holds for $J_0<2$ and contains no logarithm. At $J_0=2$ the frequency derivative that defines $\vev{E_1K_2}$ is in general discontinuous at zero frequency: its values on the two sides differ from \eqref{eq:EKidentity} by the integral of a known tail, with opposite signs. The frequency-resolved correlator at strong coupling, where $J_0=2$, is known in closed form and shows this discontinuity explicitly (appendix~\ref{app:identity}).

The form \eqref{eq:EKidentity} does not rest on the kernels. Let $\mathcal W(\omega_1,\omega_2;q)$ be the correlator of $E_{\omega_1}(n_1)$ and $E_{\omega_2}(n_2)$ in the basis of section~\ref{sec:setup}, divided by the norm, with the momentum $q$ of the source in the bra and $q+\omega_1n_1+\omega_2n_2$ in the ket, and let $\tilde{\mathcal W}$ be the same with $q\mp(\omega_1n_1+\omega_2n_2)/2$ in the bra and the ket. Hermitian conjugation of the two detectors, which commute at separated directions, makes $\tilde{\mathcal W}$ at opposite frequencies the complex conjugate of $\tilde{\mathcal W}$. With a window even in the time, $\chi(-x)=\chi(x)$, and a source of dimension two, the shift of $q$ by half the frequencies then gives the relations
\begin{align}
\vev{E_1K_2}&=\vev{E_1K_2}_{\rm sym}-\frac i2\,n_2\cdot\partial_q\vev{E_1E_2},\label{eq:symsplitE}\\
\vev{K_1K_2}&=\vev{K_1K_2}_{\rm sym}-\frac i2\big(n_1\cdot\partial_q\vev{E_1K_2}_{\rm sym}+n_2\cdot\partial_q\vev{K_1E_2}_{\rm sym}\big)\nonumber\\
&\quad-\frac14\,(n_1\cdot\partial_q)(n_2\cdot\partial_q)\vev{E_1E_2},\label{eq:symsplitK}
\end{align}
where the subscript ``sym'' marks a correlator computed with $\tilde{\mathcal W}$ in place of $\mathcal W$, that is, with the source momentum placed symmetrically in the bra and the ket. By the conjugation relation above, these correlators are real. In a conformal theory $\mathcal W$ is real \cite{KSZ2022}, so $\tilde{\mathcal W}$ is real as well and, by the conjugation relation above, unchanged when both frequencies change sign. Hence $\vev{E_1K_2}_{\rm sym}=0$, and the first relation is \eqref{eq:EKidentity}, which thus holds in any conformal theory with $J_0<2$. The last term of \eqref{eq:symsplitK}, with two derivatives of $\vev{E_1E_2}$, has no logarithm, since the energy correlator is infrared safe. For a source of another dimension, each $n_i\cdot\partial_q$ also acts on the norm, which is proportional to $(-q^2)^{\Delta-2}$. This adds $-i(\Delta-2)\vev{E_1E_2}$ to the first relation and terms of the same kind to the second, and changes none of the arguments below.

At order $a$ in $\Nfour$~SYM, on the three-parton phase space, applying \eqref{eq:Konparton} gives $\vev{E_1K_2}$ the part $-2i\vev{E_1E_2}$, and the endpoint term of section~\ref{sec:setup}, $-ia/[32\pi^2z(1-z)]$, supplies the rest of \eqref{eq:EKidentity}. Both come from the derivative term of \eqref{eq:symsplitE}. With the momentum of the bra fixed, $n_2\cdot\partial_q$ acts on the delta function of momentum as $-\partial/\partial E_2$ for the parton at $n_2$, so an integration by parts in $E_2$ splits $-(i/2)\,n_2\cdot\partial_q\vev{E_1E_2}$ into a bulk term and a boundary term at $E_2=0$. On tree amplitudes, which depend on the energies through real factors, and for a source of dimension two, whose norm does not depend on $q$, the bulk term is the value from \eqref{eq:Konparton}, and the boundary term is $-i/2$ times the boundary value at $E_2=0$, which is the endpoint term.

At leading soft power, a logarithm in $\vev{E_1K_2}$ needs $E_1$ on a hard parton and $K_2$ on a soft one or through its pair term. With $E_1$ on a soft parton, it carries a power of the soft scale $\lambda$, so the integral over $\lambda$ converges. With both detectors on hard partons, the detected energies are of order $M$, so no energy integral runs between $1/T$ and $M$, as a logarithm of the window would need. In a conformal theory with $J_0<2$, $\vev{E_1K_2}_{\rm sym}$ vanishes, so \eqref{eq:symsplitE} reduces to $\vev{E_1K_2}=-\tfrac i2\,n_2\cdot\partial_q\vev{E_1E_2}$. Since the energy correlator is infrared safe, this contains no logarithm of the window. In $\Nfour$~SYM at order $a^2$, for instance, the density of soft quanta at the detector depends on their energy, through emission from harder gluons, so the endpoint term and the action of $K$ in \eqref{eq:Konparton} on the detected gluon each carry a logarithm of the window. The two logarithms cancel, as they must, since by the relation above $\vev{E_1K_2}$ has none. By \eqref{eq:hsidentity}, which also holds with $K_2$ acting through its pair term, the same is true for $\vev{K_1K_2}$: its hard--soft contribution and the contribution with $K_2$ through its pair term have no logarithm in their sum.

\paragraph{The real part.} For the real part of $\vev{K_1K_2}$ we need neither conformal symmetry nor a particular matter content: the argument uses Hermitian conjugation and soft factorization. The hard--soft region contributes nothing to $\vev{K_1K_2}_{\rm sym}$ through order $a^2$ at leading soft power. First, the momentum constraint and the soft factor do not depend on the frequency $\omega_1$ of the detector on the hard parton. Second, the hard function here is the product of the amplitudes for the source to produce the hard partons, with the soft factor of section~\ref{sec:setup} removed, in the bra and the ket, at the energies $\bar E\mp\omega_1/2$ of the detected hard parton. It depends on the hard energy only through a real factor, apart from its phase at one loop. Third, the weight of the detector on this parton is even in $\omega_1$ for every species. Finally, the phase of the one-loop hard function is odd in $\omega_1$, but it multiplies the soft sector at order $a$, which is even in $\omega_2$. The region's contribution to $\vev{K_1K_2}$, the endpoint term of section~\ref{sec:setup} included, is therefore in the derivative terms of \eqref{eq:symsplitK}. Of these, the terms with $\vev{E_1K_2}_{\rm sym}$ and $\vev{K_1E_2}_{\rm sym}$ are imaginary, since the values taken with $\tilde{\mathcal W}$ are real, and the term with two derivatives of $\vev{E_1E_2}$ has no logarithm, so the region gives no logarithm in the real part of $\vev{K_1K_2}$. The detector's pair terms do not change this argument: with the other detector on a hard parton, where $K$ is $-i\Delta/2$, their share of $\vev{K}$ is real by Hermiticity, so the product is imaginary and they contribute only to the imaginary part of $\vev{K_1K_2}$.

In $\Nfour$~SYM at order $a$ and $z<1$, the whole correlator is the last term of \eqref{eq:symsplitK}. The terms with $\vev{E_1K_2}_{\rm sym}$ and $\vev{K_1E_2}_{\rm sym}$ vanish in a conformal theory, as shown above, and $\vev{K_1K_2}_{\rm sym}$ vanishes at this order. With the energies $\bar E_i\mp\omega_i/2$ of the detected partons in the bra and the ket, the undetected momentum in $\tilde{\mathcal W}$ does not depend on the frequencies, and the three-parton amplitudes of the $\mathbf{20'}$ source are products of powers of the parton energies at fixed directions. So $\tilde{\mathcal W}$ is even in each frequency, and its term in $\omega_1\omega_2$, which gives $\vev{K_1K_2}_{\rm sym}$, vanishes. With the one-loop energy correlator, the last term of \eqref{eq:symsplitK} gives
\begin{equation}\label{eq:KKordera}
\vev{K_1K_2}=\frac{a\,[(1+z)\ln(1-z)-2z]}{16\pi^2z^2(1-z)} .
\end{equation}

\paragraph{The imaginary part.} The imaginary part of $\vev{K_1K_2}$ vanishes in a conformal theory, where $\mathcal W$ is real. In nonconformal theories we have not shown this, and a computation with an infrared regulator does not settle it. On regulated amplitudes, applying \eqref{eq:Konparton} to the phase of the one-loop soft current \cite{CG2000b} gives an imaginary logarithm, since that phase depends on the energy relative to the scale of the regulator, but this alone does not show that the correlator has one. At order $a$ the same step, applied to the phase of the one-loop hard function, gives $\vev{K[1]}$ a real part, and at that order all other amplitudes are real, so no other contribution, the detector's pair terms included, can cancel it. The one-point function of a conformal theory \cite{KSZ2022} excludes such a real part. These two statements do not contradict each other. The reality that excludes a real part of $\vev{K[1]}$ follows from CRT acting on a null plane, and it applies at null infinity only because a conformal transformation maps null infinity to such a plane. A theory with an infrared regulator is not conformal, so its amplitudes need not respect this reality, and a regulated computation of the imaginary part is not conclusive.

\paragraph{Hard--hard region.} Here both detectors act on hard partons, and the region gives no logarithm of the window. We show this with $\tilde{\mathcal W}$, as for the real part. Since the two partons at lowest order are back to back, they reach both detectors only at $z=1$. At $z<1$ the region needs an undetected third parton, so it starts at order $a$. In $\tilde{\mathcal W}$ the frequencies shift only the energies of the two detected partons, which are of order $M$, and the total momentum of the undetected partons does not depend on them. At order $a$ the amplitudes are at tree level and depend on the energies only through real factors, so $\tilde{\mathcal W}$ is real and, by the conjugation relation, unchanged when both frequencies change sign. It therefore has no term of first order in the frequencies: $\vev{E_1K_2}_{\rm sym}$ and $\vev{K_1E_2}_{\rm sym}$ get nothing from the region at this order, and what remains to check is $\vev{K_1K_2}_{\rm sym}$, given by the term of $\tilde{\mathcal W}$ in $\omega_1\omega_2$.

At order $a^2$ the one-loop amplitude of three partons and the emission of a fourth parton have infrared divergences that cancel in their sum. The infrared poles of the one-loop amplitude depend on the energies only through a sum over pairs of $\ln s_{ij}$, with $s_{ij}=-2p_i\cdot p_j\propto E_iE_j$, and their imaginary part is proportional to the sum of the Casimirs of the partons \cite{Catani1998}. The terms of first order in the frequencies come only from the phase of the one-loop amplitude, since the emission is again at tree level, and the poles of this phase do not depend on the energies, so these terms are finite. In $\vev{K_1K_2}_{\rm sym}$ the frequencies do not spoil the cancellation either. Since $\ln s_{ij}$ is $\ln E_i+\ln E_j$ plus a term that depends only on the directions, the poles have no mixed second difference in the two frequencies, so in $\vev{K_1K_2}_{\rm sym}$ they multiply only the mixed second difference of the tree-level hard function. In every soft or collinear limit of an undetected parton the frequencies stay on hard partons: an eikonal factor does not depend on the hard energies, and a splitting collinear with a detected parton carries the frequency to its parent. The region's share of $\vev{K_1K_2}_{\rm sym}$ is then an energy correlator of the same partons, with the tree-level hard function replaced by its mixed second difference. Like $\vev{E_1E_2}$, this correlator is infrared and collinear safe, and $M$ is its only energy scale.

At $z=1$ the region starts at lowest order, with the two partons made by the source, and through order $a^2$ it also contains their one- and two-loop amplitudes. It adds no logarithm there either: the frequencies again shift only the energies of the two detected partons, and since these are of order $M$, the frequencies do not cut off any of the energy integrals. Through two loops the poles of the logarithm of the two-parton amplitude depend on the energies only through a term linear in $\ln s_{ab}$, with $s_{ab}=-2p_a\cdot p_b\propto E_aE_b$ and a real coefficient \cite{Catani1998}. Since $\ln s_{ab}$ is $\ln E_a+\ln E_b$ up to a constant, the poles have no mixed second difference, and since the coefficient is real, their imaginary part does not depend on the energies. So in the expansion of $\tilde{\mathcal W}$ in the frequencies, the terms of first order and the term in $\omega_1\omega_2$, divided by the squared amplitude, are finite. The poles of the squared amplitude multiply these terms, which vanish at tree level, and they cancel against the emission of undetected soft and collinear partons. Through its recoil, such an emission moves the two partons off the back-to-back configuration by an angle, which vanishes when the emitted partons become soft or collinear with them. The cancellation therefore takes place in the distribution in $z$ near $z=1$, as for $\vev{E_1E_2}$.

\paragraph{Pair term, other detector on a soft parton.} A pair term with the other detector on a soft parton gives no logarithm. At order $a^2$ and leading soft power the detected parton is a gluon. Let the pair term at $n_1$ absorb two partons of total energy $\omega\lesssim1/T$, and let $K$ act on the gluon, of energy $E_2$ at $n_2$. A logarithm of the window could only come from $1/T\ll E_2\ll M$. There the amplitude of this gluon in the bra is proportional to $1/E_2$, so the integrand in $\ln E_2$ is a total derivative, $-i\,dP/d\ln E_2$, where $P$ is the same integrand with $1$ in place of $K$.

A total derivative gives a logarithm only if $P$ keeps changing at large $E_2/\omega$, and we show that it tends to a constant. Since $\omega\lesssim1/T\ll E_2$, the pair is softer than every other parton, and at leading power in $\omega/E_2$ it is emitted by the eikonal current of the two hard partons and the gluon, which does not depend on $E_2$. The detector is a color singlet, so its pair term absorbs only a pair in a color-singlet state. A matter pair emitted by this current comes from one gluon and is in the adjoint, so the pair term cannot absorb it. A matter pair that emits the detected gluon can be a singlet, but it is power suppressed. The pair and the gluon then come from one virtual gluon, whose propagator $1/(2\omega E_2n_{12})$ takes the place of the factor $1/\omega^2$ of a gluon pair. The hard partons emit this virtual gluon at most as strongly as they emit the detected gluon when the pair is two gluons, up to a factor of order one, since the momentum of the pair can only increase the denominator of each eikonal factor. So this channel is suppressed by $\omega/E_2$ at every angle. The suppression survives the integral over the energy fraction of the pair, where the detector's weight makes the integrand a polynomial, and after the average over the hard axis the channel is of order $(\omega/E_2)\ln(E_2/\omega)$. At leading power the pair is then two gluons, the amplitude of the gluon in the ket is also proportional to $1/E_2$, and $P$ tends to a constant $P_\infty$ with corrections of order $\omega/E_2$.

The limit is not uniform in the angle of the gluon: near a hard parton, at a small angle $\theta$, the ratio that decides which of the two emissions comes first is $\omega/(E_2\theta^2)$ rather than $\omega/E_2$, and it is not small when $\theta^2\lesssim\omega/E_2$. There $P$ grows like $1/\theta^2$, and the two orders in which that parton emits the gluon and the pair have the weights $E_2\theta^2/(E_2\theta^2+\omega)$ and $\omega/(E_2\theta^2+\omega)$, up to factors of order one. If the pair is emitted first, it couples to the charge of the parton before the gluon is emitted, which is the sum of the charges of the parton and the gluon after it. The two orders then differ only by the gluon's color charge times the difference between the eikonal factors for emitting the pair from the gluon and from the parton, which is of order $\theta$, since their directions differ by $\theta$. So $P$ differs from its limit $P_\infty$ by an amount of order $(1/\theta)\,\omega/(E_2\theta^2+\omega)$. After the integration over $E_2$ up to the source energy $M$, the average of $P-P_\infty$ over the hard axis is of order $(\omega/M)^{1/2}$, which vanishes for a long window.

\paragraph{Crosstalk.} The crosstalk, the exchange of a scalar at zero energy, gives no logarithm of the window either. The scalar enters and leaves the detectors with energies of the order of the frequencies, which the smooth window cuts off at $1/T$, so a logarithm of the window, $\ln(MT)$, would need a dependence on $M$ from the rest of the term. The rest of the term does not depend on $M$, whichever source supplies the scalar. A scalar much softer than every other parton is emitted only at subleading power. A scalar of a soft pair with comparable energies is emitted with a soft factor that does not depend on the hard energies. Finally, at order $a^2$ a real gluon can be emitted with a color-singlet scalar pair: the energy integral then converges at large energy, and its infrared divergence cancels that of the pair's one-loop amplitude. That amplitude cannot depend on $M$. The eikonal vertices are invariant under a rescaling of the hard momenta, so $M$ could appear only through a regulator that breaks this invariance, such as a rapidity regulator. By power counting, the loop integral converges at large momentum, also where one light-cone component alone is large, and needs no such regulator.

\subsection{Angular limits and a theory without gauge fields}
\label{sec:limits}

In both limits of the angle, the results can be checked against quantities known independently of our computation: at small angle the splitting functions, and back to back an amplitude of a soft quark pair. A theory without gauge fields serves as a control, in which the soft pair scales differently with the energies and the logarithm is absent.

\paragraph{Small angle.} As $z\to0$ the pair comes from a collinear splitting of a soft gluon, so the species weights can be read off from the splitting amplitudes of a gluon, with no input from $\Nfour$~SYM:
\begin{equation}
\frac{\vev{G}}{z\ln(1/z)}\to\frac{n_s}{15}+\frac{n_f}{10}+\frac45,
\end{equation}
which is $34/15-n_f/30$ on the conformal line. The pair rates of the species are proportional, with one common factor, to $P_{g\to gg}/2=C_A(1-u)^2/u$, $P_{g\to q\bar q}=T_r(1-2u)$ per Dirac fermion and $P_{g\to s\bar s}=T_ru$ per complex scalar, with $u=x(1-x)$ and $x$ the energy fraction. In $\Nfour$~SYM, including the constant, $\vev{K_1K_2}_{\ln}\to-a^2[\ln(1/z)+1]\ln\Lambda/(16\pi^2z)$ as $z\to0$.

\paragraph{Back to back.} As $z\to1$, \eqref{eq:Gclosed} gives a finite limit, $\vev{G}\to(n_s+n_f+2)\pi^2/64-(2n_s+n_f+4)/24$, which is $\pi^2/4-1$ in $\Nfour$~SYM. At large rapidity separation the square of our amplitude of a soft fermion pair, $|s_{\rm f}|^2/\lambda^4$, reproduces the form for the soft quark pair in \cite{Monni2025}, up to an overall normalization. That work states that the divergence is power suppressed in $1-z$ against the leading back-to-back peak, and this also holds for the charge function above, since $\vev{U_f}$ tends to a constant. In QCD the leading peak has been resummed to N$^4$LL accuracy \cite{Monni2025}, and recomputed at two loops from the light transform of the correlator of four currents \cite{ChenQQC2026}.

\paragraph{A theory without gauge fields.} In $\phi^4$ theory, with the interaction $\lambda_4\phi^4/4!$ and the source $\phi^2$, the four-parton amplitude at order $\lambda_4$ is $F=\lambda_4\sum_i1/s_i$, with $s_i$ the squared invariant mass of the three partons other than $i$. Its soft pair has degree $-1$ in the energies, not $-2$, so the scale integral becomes $\int d\lambda\,\lambda$ instead of $\int d\lambda/\lambda$, and converges.

$F$ is large only where three of the partons have a small invariant mass, and on each term the action of $K[f]$ has a closed form. Take, for instance, the term $1/s_4$, with $s_4=s_{12}+s_{13}+s_{23}$, $s_{ij}=-(k_i+k_j)^2$ and $f_i=f(n_i)$. Then, by \eqref{eq:Konparton},
\begin{equation}\label{eq:Kphi4}
K[f]\,\frac{1}{s_4}=-i\,\frac{\bar f+f_4}{s_4},\qquad \bar f=\frac{f_1s_{23}+f_2s_{13}+f_3s_{12}}{s_4},
\end{equation}
where $\bar f$ is a weighted mean of $f$ over the other three partons. As the three partons become collinear, $\bar f$ tends to $f_P=f(n_P)$, with $n_P$ their common direction, so $K[f]F\to-i(f_P+f_4)F$. On the two partons made by the source, $K[f]$ multiplies the amplitude by $-i(f_1+f_2)$, and in the collinear limit it acts on $F$ in the same way, with the three partons in place of one: the factor $f_P+f_4$ depends on them only through their common direction. The collinear logarithm of the norm of this part therefore cancels against the virtual correction to the two-parton state, by unitarity, as for a collinear-safe observable. What remains of $K[f]F$ is $-i(\bar f-f_P)F$, up to the terms $1/s_i$ with $i\neq4$, which stay finite in the limit. The logarithm of the collinear cut comes from three partons at a common small angle $\theta$, where this remainder is of order $\theta$. It therefore adds no logarithm to the norm, either on its own or through its interference with $-i(f_P+f_4)F$. So $K$ has no logarithm in $\phi^4$ theory at order $\lambda_4^2$.

The absence of the logarithm in $\phi^4$ theory fits the convergence condition of \cite{KKSZ2020a}, by which $\vev{K_1K_2}$ converges absolutely in the Regge limit if $J_0<1$: the Regge intercept of this theory is of order $\lambda_4$ \cite{CaronHuot2023}, far below the marginal value $J_0=1$. It also fits an argument from spin counting, by which only two gluons reach $J_0=1$ at this order: two partons of spins $J_1$ and $J_2$ give $t$-channel spin $J_1+J_2-1$, so two gluons give spin 1 \cite{Low1975,Nussinov1975}, a gluon and a fermion give $\tfrac12$, and every other pair gives a lower spin.

\FloatBarrier

\section{The two-gluon Regge term in \texorpdfstring{$\Nfour$}{N = 4}~SYM}
\label{sec:regge}

In $\Nfour$~SYM the four-point function of $\mathbf{20'}$ operators is known at two loops and contains every contribution to $\vev{K_1K_2}$ at order $a^2$. We recover the logarithm from the Regge limit of the four-point function: the two-gluon Regge term, which has spin 1, is a power-law tail of the Mellin amplitude, and the detector kernels turn this tail into the logarithm of the window. The coefficient equals the soft-pair coefficient of section~\ref{sec:twoloop} at $(n_s,n_f)=(6,8)$, which confirms that in $\Nfour$~SYM the soft pair gives the whole logarithm. The same tail also gives a relation between this logarithm and the one in the charge correlator.

\subsection{The Regge term}
\label{sec:reggeterm}

For $\mathbf{20'}$ operators, the loop corrections to the four-point function are governed by one function \cite{EPSS2001,Eden2012}, written with the normalization of \cite{KSZ2022} as
\begin{equation}\label{eq:Mellin}
\Phi(U,V)=\int\!\frac{dj_1dj_2}{(2\pi i)^2}\big[\Gamma(1-j_1)\Gamma(1-j_2)\Gamma(1+t)\big]^2\calM(j_1,j_2)\,U^{-1-t}V^{j_2},
\end{equation}
with the Mellin variables $t=j_1+j_2$ and $\tau=(j_1-j_2)/(2i)$, so that $j_{1,2}=t/2\pm i\tau$. Here $\calM=M_{\rm KSZ}/t^2$, with $M_{\rm KSZ}$ the Mellin amplitude of \cite{KSZ2022}, and we have used the crossing symmetry $\Phi(U,V)=U^{-1}\Phi(1/U,V/U)$ to replace the factor $U^{j_1}$ of their representation by $U^{-1-t}$. The detectors sit at points~1 and~2, and the source and its conjugate at points~3 and~4. The cross ratios are $U=x_{12}^2x_{34}^2/(x_{13}^2x_{24}^2)$ and $V=x_{14}^2x_{23}^2/(x_{13}^2x_{24}^2)$. To fix the normalization of $\Phi$, take two of the three complex scalars, $Z$ and $W$, and put $\mathrm{tr}\,Z^2$ and its conjugate at points~1 and~2, and $\mathrm{tr}\,W^2$ and its conjugate at points~3 and~4, each normalized to a unit two-point function. This correlator, multiplied by $x_{12}^4x_{34}^4$, is then $1+(2/c_T)U\Phi$, with $c_T=(N^2-1)/4$: the 1 is its disconnected part, and $\Phi$ starts at one loop.

In Mellin space the Regge limit is governed by large $\tau$ at fixed $t$ \cite{CGP2012}, and a Regge term of spin 1 is a power-law tail of the amplitude, $\calM\to m(t)|\tau|^{-3}$, where the Regge residue $m(t)$ depends only on $t$. We fix $m(t)$ from the Regge limit below, and appendix~\ref{app:tail} derives this tail from the two-loop amplitude, with that $m(t)$ and with no logarithm at order $|\tau|^{-3}$. Given the tail, continue to the second sheet by $V\to Ve^{-2\pi i}$ and set $U=\sigma^2$, $1-V\simeq\sigma w_R$ with $w_R=2\cosh\rho$. Then, in the limit $\sigma\to0$,
\begin{equation}\label{eq:ftransform}
U\Phi\to \mathcal{R}(w_R)=\dint{t}\,2\pi\,\Gamma(1+t)^2\Gamma(-2t)\,m(t)\,w_R^{2t}.
\end{equation}

At two loops, in terms of the one-loop box function $\bar\Phi$, normalized in appendix~\ref{app:tail},
\begin{equation}
\Phi=-\frac a4\bar\Phi+a^2\Big[\frac{1+U+V}{32}\bar\Phi^2+\text{ladder terms}\Big].
\end{equation}
On the second sheet the box changes by its monodromy around $V=0$, and this change has a closed form at any $\sigma$, not only in the Regge limit: with $2\cosh\rho=(1+U-V)/\sigma$, the monodromy of the dilogarithm gives $\delta\bar\Phi=2\pi i\rho/(\sigma\sinh\rho)$. This grows like $1/\sigma$, so the term in $\bar\Phi^2$ grows like $1/\sigma^2$. Since $U=\sigma^2$, the contribution of this term to $U\Phi$ then tends to a nonzero function of $w_R$ as $\sigma\to0$, the behavior of a Regge term of spin 1. So at two loops the Regge function $\mathcal{R}(w_R)$ of \eqref{eq:ftransform}, the limit of $U\Phi$, is
\begin{equation}\label{eq:f2}
\mathcal{R}(w_R)=-\frac{\pi^2}{4}\,a^2\,\beta(w_R)^2,\qquad\beta(w_R)=\frac{\rho}{\sinh\rho}.
\end{equation}
For $w_R<2$ the variable $\rho$ is imaginary. The ladder terms cannot reach spin 1: on the second sheet they grow only like $\ln^2\sigma/\sigma$, against $\sigma^{-2}$ for $\bar\Phi^2$, so their part of $U\Phi$ vanishes in the Regge limit (appendix~\ref{app:tail}). In Mellin space, the amplitude of $\bar\Phi^2$ is a convolution of two one-loop amplitudes. The exact monodromy reduces it to a onefold Mellin--Barnes integral and a remainder of relative order $\ln\tau/\tau$, and the first pole of that integral gives the Mellin transform of $\beta(w_R)^2$, and hence of $\mathcal{R}(w_R)$.

The function $\mathcal{R}(w_R)$ is not new. It is two-gluon exchange in a color singlet, a universal pomeron propagator $\tanh(\pi\nu/2)/\nu$ with impact factors \cite{CCP2008}. In the form quoted in \cite{CGP2012}, with $g_{\rm CGP}^2=a/4$ in our normalization,
\begin{equation}\label{eq:alpha}
\alpha(\nu)=i\,\frac{16\pi^5g_{\rm CGP}^4}{N^2}\,\frac{\tanh(\pi\nu/2)}{\nu\cosh^2(\pi\nu/2)},
\end{equation}
and the Regge form given there at spin 1 is $2\pi i\int d\nu\,\alpha(\nu)\Omega_{i\nu}(\rho)$, with $\Omega_{i\nu}=\nu\sin\nu\rho/(4\pi^2\sinh\rho)$. This equals $-(2\pi^2a^2/N^2)\beta^2$, which is $(2/c_T)\mathcal{R}(w_R)$ at large $N$, sign and factor included.

In the Mellin variable $t$, where the detector kernels are simple, the residue has a closed form. Inverting \eqref{eq:ftransform} gives, for $-1<\mathrm{Re}\,t<0$,
\begin{equation}\label{eq:mclosed}
m(t)=\frac{1}{\pi\Gamma(1+t)^2\Gamma(-2t)}\int_0^\infty\!\frac{dw_R}{w_R}w_R^{-2t}\mathcal{R}(w_R)=-\frac\pi2\,a^2\int_0^{\pi/2}\!d\phi\,\phi\,\sin^{2t}\!\phi .
\end{equation}
To obtain the last form, write $\beta(w_R)=\int_0^\infty d\kappa/(\kappa^2+w_R\kappa+1)$, combine the two denominators of $\beta^2$ with a Feynman parameter, do the $w_R$ integral, and read the remaining double integral as one over a disk in polar coordinates about a point on its boundary.

\subsection{The logarithm from the Regge term}
\label{sec:identity}

We write $\vev{E_{\omega_1}(n_1)E_{\omega_2}(n_2)}=(1/8\pi^2)\mathrm{EEC}(2\omega_1,2\omega_2,z)$, with $\partial_1$ and $\partial_2$ the derivatives in its first two arguments. Expanded at small frequency, the exact kernels \cite{KSZ2022} have a coefficient $c_{mn}$ at order $m$ in $2\omega_1$ and $n$ in $2\omega_2$, and they give
\begin{equation}\label{eq:kernel}
\partial_1\partial_2\mathrm{EEC}\big|_{\omega_{1,2}=0}=\dint{t}\int\!\frac{d\tau}{2\pi}\,\frac{\pi}{2\sin\pi t}\,z^{-t}(1-z)^{t-1}\,t^2\calM\,c_{11},
\end{equation}
with $t^2c_{11}\to\tau^2[t^2+(1-z)(1+t)]/z^2$ at large $\tau$. The same expression with $c_{00}=1/z^2$ in place of $c_{11}$ gives EEC itself, and there the $\tau$ integral converges. We fix the prefactor by the one-loop correlator: on $\calM=-(a/4)/(j_1j_2)^2$ it gives $\mathrm{EEC}=-a\ln(1-z)/[4z^2(1-z)]$, as it must \cite{BHKSZ2014c}. Since $t^2c_{11}$ grows like $\tau^2$ at large $\tau$ and $c_{00}$ does not, a tail $\calM\sim|\tau|^{-3}$ gives a logarithm for $K$ and none for $E$, through
\begin{equation}\label{eq:tauint}
\int\!\frac{d\tau}{2\pi}\,\tau^2|\tau|^{-3}=\frac1\pi\ln\Lambda .
\end{equation}
This is the marginal case of the convergence condition quoted in section~\ref{sec:setup}, by which $\vev{KK}$ needs $J_0<1$: at $J_0=1$ the tail is $|\tau|^{-3}$, so the $\tau$ integral diverges logarithmically, and \eqref{eq:tauint} gives the coefficient of the logarithm.

At nonzero frequencies the $\tau$ integral is cut off at large $|\tau|$. In that region the exact kernels depend on the frequencies only through the products $\omega_1\tau$ and $\omega_2\tau$, for either sign of the frequencies (appendix~\ref{app:identity}), so the expansion in the frequencies holds only for $|\tau|\ll M/\omega_{1,2}$, and with $\omega_{1,2}\sim1/T$ the cutoff is $MT$ up to a constant. With one frequency, the exact energy kernel \cite{KSZ2022} is its zero-frequency form times
\begin{equation}\label{eq:onefreq}
\Big(1+\frac{2\omega_2}{M}\Big)^{-j_1}\Big(1+\frac{2\omega_2}{M}+\frac{2\omega_2^2}{3M^2}\Big),
\end{equation}
which at large $\tau$, with $\tau\omega_2$ fixed, is the phase $e^{-2i\tau\omega_2/M}$ of an arrival time $u_2=-2\tau/M$, and the other detector sees $u_1=2\tau/M$, so the Mellin variable $\tau$ is a detector time in units of $2/M$. The leading behavior of the energy correlator at small frequency is related to the Regge limit of the Mellin amplitude \cite{KSZ2022}.

To obtain the coefficient of the logarithm, we define the function $\Psi$, the transform of the residue $m(t)$ with the kernel $X^t\pi/\sin\pi t$,
\begin{equation}\label{eq:Psi}
\Psi(z)=\frac1\pi\dint{t}\frac{\pi}{\sin\pi t}X^tm(t)=\frac{a^2}{2}\,z\,\calI(z),\qquad X=\frac{1-z}{z},
\end{equation}
where the second equality uses \eqref{eq:mclosed} and $\int dt/(2\pi i)\,(\pi/\sin\pi t)\,\xi^t=-1/(1+\xi)$ for $-1<\mathrm{Re}\,t<0$, at $\xi=X\sin^2\phi$. The factor $t$ in the kernel acts as $X\partial_X=(1-z)\partial_\eta$ on $X^t$, with $\eta=\ln(1/z)$. Then, with $\vev{K_1K_2}=-(1/2\pi^2)\partial_1\partial_2\mathrm{EEC}|_{\omega_{1,2}=0}$,
\begin{equation}\label{eq:operator}
z^2\vev{K_1K_2}_{\ln}=-\frac{\ln\Lambda}{4\pi^2}\,\calO\Psi(z),\qquad\calO=(1-z)\partial_\eta^2+\partial_\eta+1 .
\end{equation}
The operator acts simply on this $\Psi$, since the derivative of $\sqrt z\,\calI(z)$ closes on elementary functions, $\partial_\eta[\sqrt z\,\calI(z)]=\sqrt z\ln z/[4(1-z)]$. This gives
\begin{equation}
\calO\big[\tfrac12z\,\calI(z)\big]=\frac{z(3-z)}{8}\,\calI(z)-\frac z8\big(1+\tfrac12\ln z\big),
\end{equation}
and so, for $\Nfour$~SYM at two loops,
\begin{equation}\label{eq:KKclosed}
\vev{K_1K_2}_{\ln}=-\frac{a^2}{32\pi^2z}\Big[(3-z)\calI(z)-1-\tfrac12\ln z\Big]\ln\Lambda .
\end{equation}

This equals the soft-pair result \eqref{eq:KKgeneral}, normalization and elementary terms included. The closed form \eqref{eq:Gclosed} of the species sum gives $\vev{G}=z[(3-z)\calI-1-\tfrac12\ln z]$ at $(n_s,n_f)=(6,8)$ with no input from the correlator. The agreement is also visible in the integrands over the energy asymmetry $w$: in the integral form \eqref{eq:KKw} they differ by a total derivative,
\begin{equation}\label{eq:totalder}
\frac z8\Big[(1+v^2)\frac{\artanh v}{v}-1\Big]-\calO\Big[\frac z4\,\frac{\artanh v}{v}\Big]=\frac18\,\partial_w\Big[\tilde p\,\frac{\artanh v}{v}+\tilde q\Big],
\end{equation}
with $\tilde p=z^2w(w^2-1)(1+v^2)/(2v^2)$ and $\tilde q=z^2w(1-w^2)/(2v^2)$, and the boundary terms vanish since $(1-w^2)\artanh v\to0$ at $w=\pm1$. So the soft-pair logarithm and the spin-1 term of the correlator are one function, with no free constant.

The correlator side takes no input from section~\ref{sec:twoloop}, and the four-point function contains every contribution at order $a^2$. Hence in $\Nfour$~SYM the agreement between the Regge term and the soft pair also shows that the other contributions together give no logarithm, although it does not show that each of them gives none by itself, and it says nothing about other matter content. At order $a^2$ the soft-pair logarithm is thus the logarithm that the convergence condition of \cite{KKSZ2020a} allows at the marginal Regge intercept $J_0=1$.

\subsection{Relation to the charge correlator}
\label{sec:QQrelation}

The charge correlator sees the same spin-1 tail of the Mellin amplitude, $m(t)|\tau|^{-3}$, through a simpler kernel. Exact relations between the kernels of scalar, charge and energy detectors are known in the top $SU(4)$ channel \cite{KSZ2022}; the charge correlator here is not the one in that channel, and its kernel differs from the energy kernel by a factor that depends on the Mellin variables. As in section~\ref{sec:general}, the hard partons are neutral under $Q$. We combine the channel kernels of \cite{BHKSZ2014b} with the $R$-symmetry factors given there and normalize by the direct one-loop rate. Then
\begin{equation}\label{eq:FQ}
\vev{Q_1Q_2}=\frac{F_Q}{16\pi^2z^2},\qquad F_Q=-\trQ\dint{t}\int\!\frac{d\tau}{2\pi}\,\frac{4\pi}{\sin\pi t}\,X^tj_1j_2\,\calM,
\end{equation}
which, with the tail $m(t)|\tau|^{-3}$, gives
\begin{equation}\label{eq:QQfromPsi}
z^2\vev{Q_1Q_2}_{\ln}=-\frac{\trQ\,\ln\Lambda}{4\pi^2}\,\Psi(z).
\end{equation}
This is \eqref{eq:QQN4}. Together with \eqref{eq:operator} it gives, for any residue $m(t)$ with $\calM\to m(t)|\tau|^{-3}$,
\begin{equation}\label{eq:operatorQQ}
z^2\vev{K_1K_2}_{\ln}=\calO\Big[\frac{z^2\vev{Q_1Q_2}_{\ln}}{\trQ}\Big].
\end{equation}
At two loops the result closes on the same function,
\begin{equation}\label{eq:KKQQ}
\vev{K_1K_2}_{\ln}=\frac{3-z}{4}\,\frac{\vev{Q_1Q_2}_{\ln}}{\trQ}+\frac{a^2}{32\pi^2z}\Big(1+\tfrac12\ln z\Big)\ln\Lambda .
\end{equation}
This is the second line of \eqref{eq:intro}. As $z\to0$ the two terms have the same sign and the ratio 3 to 1 in the coefficient of $\ln(1/z)/z$. The elementary term changes sign at $z=e^{-2}$, and at $z=1$ the factor $(3-z)/4$ is $\tfrac12$. On Regge partial waves the operator in \eqref{eq:operatorQQ} is diagonal, with eigenvalue $(3-\nu^2)/4$, as section~\ref{sec:beyond} shows.

The function $\calI(z)$ is not new either. The weight-two term with $\sqrt z$ in the two-loop energy correlator of $\Nfour$~SYM is
\begin{equation}\label{eq:F2term}
4\sqrt z\Big[\Li(-\sqrt z)-\Li(\sqrt z)+\frac{\ln z}{2}\ln\frac{1+\sqrt z}{1-\sqrt z}\Big]=-8z\,\calI(z),
\end{equation}
in the function $F_2(z)$ of \cite{BHKSZ2014c}. So the correlator $\vev{E_1E_2}$, which converges, contains the function that controls the divergence of $\vev{Q_1Q_2}$ and $\vev{K_1K_2}$. We observe this coincidence but do not attempt to explain it here.

\section{The time-weighted detector as an operator}
\label{sec:operator}

Sections~\ref{sec:twoloop} and~\ref{sec:regge} found the logarithm in the product of two detectors at separated directions. This section shows what it means for the smeared detector $K[f]$ as an operator: unless $f$ is constant, the states created by the source lie outside its domain, since the squared norm of $K[f]$ acting on them diverges logarithmically in the infrared. The coefficient of this divergence follows from that of $\vev{K_1K_2}_{\ln}$, and these statements hold at order $a^2$.

\subsection{The one-body operator and its domain}
\label{sec:onebody}

At zero frequency $K[f]$ conserves the number of partons and acts on each as in \eqref{eq:Konparton}. On the wavefunction $\hat F=\sqrt E\,F$, normalized in the measure $dE\,d^2n$, it is $f(n)D$ with $D=-i(E\partial_E+\tfrac12)$, the generator of the unitary group $\hat F(E,n)\to e^{\epsilon f(n)/2}\hat F(e^{\epsilon f(n)}E,n)$. So for real bounded $f$ it is self-adjoint on the one-parton space, by Stone's theorem, and its second quantization is self-adjoint on Fock space. It annihilates the vacuum, and all the $K[f]$ commute. In the Mellin basis $D$ multiplies $E^{-1/2+i\kappa}$ by $\kappa$, so its spectrum is the real line. $K[f]$ is thus a well-defined self-adjoint operator on Fock space, and the only question is whether the states created by the source lie in its domain.

On the soft pair of section~\ref{sec:softpair}, with $F=s(y)/\lambda^2$, the two partons enter with opposite signs, $(E_1\partial_1+1)+(E_2\partial_2+1)=0$, so $K[f]F=-i(f_1-f_2)\,\partial_yF$, where $f_1$ and $f_2$ are the values of $f$ at the directions of the two partons. The group acts as $e^{iK[f]}F=e^{f_1+f_2}F(e^{f_1}E_1,e^{f_2}E_2)$, which on the pair is $s(y+f_1-f_2)/\lambda^2$: the factor $1/\lambda^2$ stays fixed and $s$ shifts. Let $\psi$ be the state created by the source and $\mathrm{KK}(n_1,n_2)$ the coefficient of $\ln\Lambda$ in $\vev{K_1K_2}$ at separated directions. Because of the opposite signs, $K$ acting on one parton of the pair is $-K$ acting on the other. A term with both detectors on the same parton therefore equals the negative of the corresponding term with the detectors on different partons, integrated over the direction of the other parton. This fixes the contact term at coincident directions, and the logarithm in the norm of $K[f]\psi$ follows:
\begin{equation}\label{eq:contact}
\begin{aligned}
\vev{K_1K_2}_{\ln}&=\Big[\mathrm{KK}(n_1,n_2)-\delta^2(n_1,n_2)\int\!d^2n'\,\mathrm{KK}(n',n_2)\Big]\ln\Lambda,\\
\|K[f]\psi\|^2_{\ln}&=\frac12\ln\Lambda\iint(f_1-f_2)^2(-\mathrm{KK}) .
\end{aligned}
\end{equation}
Each of the two terms of the first line diverges when smeared, since $\mathrm{KK}$ grows like $\ln(1/z)/z$ at small angle; only their sum is finite, and it gives the second line, where the integral is over the two directions. So in the spherical-harmonic modes $K_{\ell m}=K[Y_{\ell m}]$ the logarithm drops out of the mode $\ell=0$, and all norms are nonnegative by construction. For a normalized harmonic the norm of $K_{\ell m}$ is
\begin{equation}\label{eq:modes}
\vev{K_{\ell m}K_{\ell m}^\dagger}_{\ln}=4\pi\ln\Lambda\int_0^1\!dz\,\big[1-P_\ell(1-2z)\big](-\mathrm{KK}),
\end{equation}
and in $\Nfour$~SYM the moments $\int_0^1dz\,z^k\calI(z)=[\pi^2/2-\psi'(k+1)]/(4k+2)$ give each of these norms in closed form. The integral reaches the back-to-back point $z=1$, where $\mathrm{KK}$ has a finite limit (section~\ref{sec:limits}) and the hard--hard region adds no logarithm (section~\ref{sec:hardsoft}).

With a smooth infrared cut on the energies and a collinear cut on the angles, the state lies in Fock space at order $a^2$. The detector $K[f]$ rescales the energies at fixed directions, so it does not act on the collinear cut, and for smooth $f$ the norm in \eqref{eq:contact} is finite without that cut, since $(f_1-f_2)^2$ vanishes like $z$ as the two directions approach. As the infrared cut is removed, $\|K[f]\psi\|^2$ grows logarithmically. Whatever the form of that cut, the coefficient of the logarithm includes the term in \eqref{eq:contact}, which vanishes only for constant $f$. So in perturbation theory the states created by the source lie outside the domain of $K[f]$ for every nonconstant $f$.

The two lowest modes make this concrete. The mode $\ell=0$, with $f$ constant, is the dilatation, and the state lies in its domain. The modes $\ell=1$ show that a charge that is finite on the state can contain a detector that diverges on it. The boost along the conformal Killing vector $Y^A=\nabla^AY_{1m}$ is $N[Y]+\tfrac12K[\nabla\cdot Y]$, with $N[Y]$ the angular flux, and since $\nabla\cdot Y=-2Y_{1m}$ it equals $N[Y]-K_{1m}$. So $K_{1m}$ has a logarithm although the boost has none, and $N[Y]\psi=K_{1m}\psi$ plus a finite term. In $\Nfour$~SYM the logarithm is
\begin{equation}\label{eq:K1m}
\vev{K_{1m}K_{1m}^\dagger}_{\ln}=\Big(\frac{\pi}{9}-\frac{1}{6\pi}\Big)a^2\ln\Lambda .
\end{equation}

\subsection{The windowed detector}
\label{sec:vacuum}

The time window that cuts off the logarithm also changes the operator. At zero frequency $K[f]$ conserves the number of partons and annihilates the vacuum, as in section~\ref{sec:onebody}. The windowed detector $K_\chi[f]$, with the weight $W(u)=u\chi(u/T)$ in place of $u$, also has the pair terms $\hat a\hat a$ and $\hat a^\dagger\hat a^\dagger$ of section~\ref{sec:setup}. It therefore does not annihilate the vacuum: the term $\hat a^\dagger\hat a^\dagger$ creates two partons in the same direction, so the squared norm of $K_\chi[f]|0\rangle$ is $\delta^2(0)\int d^2n\,f^2$ times a positive integral over the energies of the pair. For a conformally coupled scalar, as in $\Nfour$~SYM, this integral itself diverges logarithmically where the energy of either parton of the pair vanishes. With the factor $\delta^2(0)$, $K_\chi[f]$ thus has infinite norm on every state at every window.

The same structure appears for frequency-resolved flow operators \cite{KZ2021}, whose vacuum two-point function is a delta function with a divergent coefficient because each parton of the pair brings one delta function. In the commutator of two local stress tensors, the term proportional to the identity operator, which is a c-number fixed by the two-point function, diverges in the light-sheet limit, where the two stress tensors lie on a common null plane. This term vanishes when the integrated operators $\int du\,T_{uu}$, $\int du\,u\,T_{uu}$ and $\int du\,T_{uA}$ are commuted with one another \cite{Besken2021}. The vacuum term sits at coincident directions, so it does not touch the correlator at separated directions. For a smeared detector, our statements about the window concern the connected part, $\langle\psi|K_\chi[f]^2|\psi\rangle-\langle0|K_\chi[f]^2|0\rangle$. For a conformally coupled scalar the subtraction of the vacuum term does not ensure the finiteness of this part: the crosstalk of section~\ref{sec:hardsoft} carries the factor $1/z$ \cite{KSZ2022}, which is not integrable at coincident directions when both detectors are smeared with $f$. Nevertheless, at leading soft power the crosstalk does not depend on $M$, so it adds no logarithm of $M$ to this part.

In a normalizable state the same pair term removes the one-gluon contribution. In position space the eikonal field of the soft gluon is a pulse at $u=0$: its field strength is proportional to $\delta(u)$, so the term of $K_\chi$ linear in it carries $W(u)\,\delta(u)$, which vanishes since $W(0)=0$. In energy space, with $E$ the energy of a mode of the eikonal field, $E'$ the outgoing energy and $\hat W$ the Fourier transform of $W$, the number-conserving and pair terms of $K_\chi$ combine into $\int_0^\infty dE\,[\hat W(E'-E)+\hat W(E'+E)]=2\pi W(0)=0$ at every $E'$. Without the pair term the integral would not vanish below $E'\sim1/T$. In the basis of section~\ref{sec:setup}, however, with a momentum eigenstate in the bra, the expectation value of $K_\chi$ still contains the endpoint term of that section, which does not depend on the window. In this basis the pair term removes only the part of the one-gluon contribution that depends on the window. So the connected part has no logarithm from one soft gluon at any window, and at order $a^2$ its dependence on $M$ is the logarithm of \eqref{eq:contact}, with $\ln\Lambda=\ln(MT)$. We quote the coefficient with the window rather than with a cut on the energies, because $K$ also acts on such a cut. With a cut function $c(E)$ on each energy, $K(cF)=c\,KF-i\,(E\partial_Ec)\,F$ on a parton of energy $E$. The second term depends on the form of the cut, and it does not vanish on a gluon at the cut, since for two gluons the function $s(y)$ in $F=s(y)/\lambda^2$ tends to a constant at large $|y|$, where the other gluon has a much larger energy.

\subsection{Wave packets and the one-point function}
\label{sec:packets}

So far we have worked in the basis of section~\ref{sec:setup}, with a source at the origin in the ket and a momentum eigenstate in the bra, as in \cite{BHKSZ2014b,BHKSZ2014c}. In this basis $K$ acts only on the ket amplitude. Since momentum eigenstates are not normalizable, we now turn to a normalizable wave packet. Unlike the energy detector $E(n)$, the time-weighted detector $K(n)$ is not translation invariant: a translation by $x$ shifts the arrival times, so that $K(n)\to K(n)-(n\cdot x)E(n)$ in the mostly plus metric. A packet $\int d^4x\,g(x)O(x)|0\rangle$ superposes the source at different points, and its amplitude at total momentum $P$ is $\tilde g(P)F$, on which $K[f]$ acts as
\begin{equation}\label{eq:packetaction}
K[f]\,(\tilde gF)=\tilde g\,K[f]F-i\,\partial_\mu\tilde g\,E[n^\mu f]F .
\end{equation}
The norm then has terms with $\vev{K[f]K[f]}$, $\vev{E[n^\mu f]K[f]}$ and $\vev{E[n^\mu f]E[n^\nu f]}$. The exact relation, with $\mathcal W$ as in section~\ref{sec:hardsoft} and $\rho$ the spectral density of the source, is
\begin{equation}\label{eq:packet}
\begin{gathered}
\langle\psi|E_{\omega_1}(n_1)E_{\omega_2}(n_2)|\psi\rangle=\int\!\frac{d^4P}{(2\pi)^4}\,\tilde g^*(P)\,\tilde g(P+\omega_1n_1+\omega_2n_2)\,\rho(P)\,\mathcal W(\omega_1,\omega_2;P),\\
\langle O(P')|O(P)\rangle=(2\pi)^4\delta^4(P-P')\,\rho(P).
\end{gathered}
\end{equation}

When the window is much longer than the extent of the packet and the displacement of its center from the origin, the frequencies are small compared with the scale on which $\tilde g$ varies. Expanded to second order in them about the midpoint $P+(\omega_1n_1+\omega_2n_2)/2$, the term of \eqref{eq:packet} in $\omega_1\omega_2$ has three parts: $\vev{K_1K_2}_{\rm sym}$ averaged over $|\tilde g|^2\rho$; $\vev{E_1K_2}_{\rm sym}$ and $\vev{K_1E_2}_{\rm sym}$ with the same average, times the gradient of the phase of $\tilde g$; and $\vev{E_1E_2}$ times terms with two derivatives of $\tilde g$. The last has no logarithm, since energy correlators are safe, and in a conformal theory the second vanishes (section~\ref{sec:hardsoft}). So on a normalizable state in a conformal theory, with such a window, the coefficient of the logarithm is the average of the momentum-eigenstate coefficient over $|\tilde g|^2\rho$, with $M$ and $z$ taken in the rest frame of each $P$. In nonconformal theories the second part vanishes when $\tilde g$ has a constant phase, and the coefficient of the logarithm on the packet is then the average over $|\tilde g|^2\rho$ of the real part of the momentum-eigenstate coefficient. A translation of the packet away from the origin, however, gives the phase a gradient.

The detectors and their windows are defined in one frame, but the coefficient at each $P$ can be computed in the rest frame of $P$. Under the boost to that frame, each direction $n$ goes to another direction $n'$. Since $K(n)$ is homogeneous of degree $-2$ in the null vector $n$, the detector at $n$ becomes the detector at $n'$ times $M^2/(P\cdot n)^2$. The boost also rescales the width $T$ of the window at $n$, by a factor that depends on the direction. This adds a constant to $\ln\Lambda=\ln(MT)$ and leaves its coefficient unchanged. In the original frame the coefficient at $P$ is therefore the coefficient for a source at rest, multiplied by $(P^2)^2(P\cdot n_1)^{-2}(P\cdot n_2)^{-2}$ and taken at $z'=-P^2z/[(P\cdot n_1)(P\cdot n_2)]$, the value of $z$ in the rest frame of $P$. In a conformal theory the coefficient thus depends on the packet only through $|\tilde g|^2$, not on its phase. In particular, if the mean momentum of the packet is at rest and the momenta spread about it by $\delta P\ll M$, the average equals the coefficient at rest up to relative corrections of order $(\delta P)^2/M^2$.

The finite part of $\vev{K_1K_2}$, the term without the logarithm, depends on the packet, in particular through the arrival times: a source at the point $x$ shifts the arrival time at the detector in the direction $n_i$ by $-n_i\cdot x$. These shifts enter the third part of the expansion, the one with $\vev{E_1E_2}$, which is the average over the packet of $\vev{E_1E_2}$ times their product. This part is therefore of order $\vev{E_1E_2}$ times the square of the extent of the packet or of the displacement of its center from the origin, whichever is larger.

In the basis of section~\ref{sec:setup} the bra and the ket are different states, so the correlators can have imaginary parts, such as the endpoint term of that section. They all come from the derivative terms of \eqref{eq:symsplitE} and \eqref{eq:symsplitK}, that is, from placing the momentum of the source asymmetrically in the bra and the ket. On a normalizable state the detectors are Hermitian and commute at separated directions, so the correlators are real, and these imaginary parts have no counterpart.

In a conformal theory the one-point function $\vev{K(n)}$ has no logarithm. Since $K[1]$ is the dilatation, it acts on $O(0)|0\rangle$ as $-i\Delta$: the amplitude for $k$ partons has degree $\Delta-k$ in the energies, so $\sum_i(E_i\partial_i+1)F=\Delta F$. With Lorentz invariance this fixes
\begin{equation}\label{eq:firstmoment}
\vev{K(n)}=-\frac{i\Delta}{4\pi}\,\frac{-q^2}{(q\cdot n)^2}
\end{equation}
in the state with momentum $q$, at any coupling, provided that the phase-space integral that defines $\vev{K(n)}$ converges. One-point functions of this kind are also computed in \cite{HP2026}. At tree level with $\Delta=2$ and a source at rest, \eqref{eq:firstmoment} gives $-2i/4\pi$, which comes from the two hard partons. So, unlike the product of two detectors, a single detector is finite in a conformal theory.

\section{Beyond two loops}
\label{sec:beyond}

The logarithm \eqref{eq:KKclosed} is a fixed-order result: at higher orders each further power of $a$ can come with a further power of $\ln\Lambda$, so it describes the dependence on the window only while $a\ln\Lambda\ll1$. In this section we carry it beyond two loops in planar $\Nfour$~SYM, in the state created by a $\mathbf{20'}$ operator. We resum the leading logarithms, which grow as a power of the window set by the BFKL intercept, find their term at three loops in closed form, and argue that the growth stays a power, set by the planar intercept, from weak to strong coupling.

\paragraph{Partial waves.} We resolve the two-loop result into Regge partial waves, because beyond two loops each wave grows with its own power of the window and the operator $\calO$ of \eqref{eq:operator} is diagonal on them. In the Regge variable $\nu$, the residue \eqref{eq:mclosed} is a superposition of Mellin partial waves,
\begin{equation}\label{eq:mnu}
m(t)=-\frac{\pi^2a^2}{16}\int_{-\infty}^\infty\!d\nu\,\tanh^2\frac{\pi\nu}{2}\;\frac{\Gamma\big(t+\frac{1+i\nu}{2}\big)\Gamma\big(t+\frac{1-i\nu}{2}\big)}{\Gamma(1+t)^2},\qquad-\tfrac12<\mathrm{Re}\,t<0 .
\end{equation}
This follows by writing \eqref{eq:f2} with the $\nu$ dependence of \eqref{eq:alpha}, so that $\beta^2$ is the integral over $\nu$ of $\pi^4\tanh(\pi\nu/2)\,\Omega_{i\nu}(\rho)/[\nu\cosh^2(\pi\nu/2)]$, and taking the Mellin transform of the harmonic function:
\begin{equation}\label{eq:mellinharm}
\int_0^\infty\!\frac{dw_R}{w_R}\,w_R^{-2t}\,\frac{\sin\nu\rho}{\sinh\rho}=\frac{\sinh\pi\nu}{2\pi}\,\Gamma(-2t)\,\Gamma\Big(t+\frac{1+i\nu}{2}\Big)\Gamma\Big(t+\frac{1-i\nu}{2}\Big).
\end{equation}
Each partial wave gives a conical function in the transform $\Psi$ of \eqref{eq:Psi}, since closing the contour on the poles of $1/\sin\pi t$ gives the hypergeometric series of $-P_{-\frac12+\frac{i\nu}{2}}(1+2X)/\cosh(\pi\nu/2)$ for $X<1$, and analytic continuation extends the equality to all $X>0$, with $1+2X=2/z-1$. So, in partial waves,
\begin{equation}\label{eq:Psinu}
\Psi(z)=\frac{\pi^2a^2}{16}\int_{-\infty}^\infty\!d\nu\,\frac{\sinh^2(\pi\nu/2)}{\cosh^3(\pi\nu/2)}\,P_{-\frac12+\frac{i\nu}{2}}\Big(\frac2z-1\Big).
\end{equation}
In the variable $x=2/z-1$ the operator of \eqref{eq:operator} reads $\calO=(x^2-1)\partial_x^2+2x\partial_x+1$, the Legendre operator plus one, so it is diagonal on these functions:
\begin{equation}\label{eq:Oeigen}
\calO\,P_{-\frac12+\frac{i\nu}{2}}\Big(\frac2z-1\Big)=\frac{3-\nu^2}{4}\,P_{-\frac12+\frac{i\nu}{2}}\Big(\frac2z-1\Big).
\end{equation}
With \eqref{eq:operator}, the logarithm in partial waves is
\begin{equation}\label{eq:KKnu}
\vev{K_1K_2}_{\ln}=-\frac{a^2\ln\Lambda}{256z^2}\int_{-\infty}^\infty\!d\nu\,(3-\nu^2)\,\frac{\sinh^2(\pi\nu/2)}{\cosh^3(\pi\nu/2)}\,P_{-\frac12+\frac{i\nu}{2}}\Big(\frac2z-1\Big),
\end{equation}
which is \eqref{eq:KKclosed} resolved into partial waves, and by \eqref{eq:QQfromPsi} the charge correlator $\vev{Q_1Q_2}_{\ln}/\trQ$ is the same integral with the eigenvalue $(3-\nu^2)/4$ replaced by 1.

\paragraph{The leading-logarithm tower.} In the planar theory the Regge term exponentiates \cite{CGP2012,CDGP2014}. Beyond two loops the Regge limit has leading logarithms $a^{2+k}\ln^k\sigma$, and in each partial wave of $\mathcal{R}(w_R)$ they sum to a power, $\sigma^{1-j(\nu)}$, with the BFKL spin \cite{Lipatov1976,KLF1977,BL1978}
\begin{equation}\label{eq:BFKL}
j(\nu)=1+a\Big[2\psi(1)-\psi\Big(\frac{1+i\nu}{2}\Big)-\psi\Big(\frac{1-i\nu}{2}\Big)\Big].
\end{equation}
In the Mellin integral \eqref{eq:Mellin}, the Regge limit \eqref{eq:ftransform} comes from $|\tau|$ of order $1/\sigma$. An extra factor $|\tau|^p$ in the tail of the amplitude therefore multiplies the Regge limit by $\sigma^{-p}$, and at leading logarithm the factor $\sigma^{1-j(\nu)}$ is the factor $|\tau|^{j(\nu)-1}$ in the tail. The same factor enters the detector kernels, where it turns the integral \eqref{eq:tauint} into that of $\tau^2|\tau|^{j(\nu)-4}$. Cut off at $|\tau|\sim MT$, where the expansion in the frequencies stops (section~\ref{sec:identity}), this integral has the derivative $\Lambda^{j(\nu)-1}/\pi$ in $\ln\Lambda$ instead of $1/\pi$, so \eqref{eq:KKnu} becomes the tower of leading logarithms
\begin{equation}\label{eq:LL}
\frac{d\vev{K_1K_2}}{d\ln\Lambda}=-\frac{a^2}{256z^2}\int_{-\infty}^\infty\!d\nu\,(3-\nu^2)\,\frac{\sinh^2(\pi\nu/2)}{\cosh^3(\pi\nu/2)}\,\Lambda^{j(\nu)-1}\,P_{-\frac12+\frac{i\nu}{2}}\Big(\frac2z-1\Big).
\end{equation}
The tower depends only on the two-loop residue resolved in $\nu$ and on the spin \eqref{eq:BFKL} at leading order. Everything else changes only subleading logarithms: the residue at order $a^3$, the spin at the next order, the dependence of the partial waves on $j$, the shape of the cutoff and the constant between $\Lambda$ and $MT$. In the four-point function, the same two functions fix the leading logarithm $a^n\ln^{n-2}\sigma$ of the Regge limit at every loop order $n$ \cite{CDGP2014}. At three loops the term $a^3\ln\sigma$ is also known directly, from the Regge limit of the three-loop correlator \cite{DDEHPS2013,CDGP2014}, so the term of order $a^3$ in the tower follows from that correlator without assuming the exponentiation.

\paragraph{Three loops.} At order $a^3$ the tower \eqref{eq:LL} gives $\vev{K_1K_2}$ the term $C_3(z)\,a^3\ln^2\Lambda$. In the form of \eqref{eq:operator}, the tower reads $z^2\,d\vev{K_1K_2}/d\ln\Lambda=-\calO\Psi/(4\pi^2)$, with each partial wave of $\Psi$ in \eqref{eq:Psinu} multiplied by $\Lambda^{j(\nu)-1}$. Expanded in $a$, this $\Psi$ is $\frac12a^2z\calI(z)+a^3z\calI_3(z)\ln\Lambda+O(a^3,\,a^4\ln^2\Lambda)$, where $O(a^3)$ stands for terms without a logarithm, which the tower does not fix. We obtain $C_3$ in three steps: the $\nu$ integral gives the residue $m_3(t)$ of the term in $\ln|\tau|$ in the tail, the $t$ integral gives $\calI_3$, and $\calO$ gives $C_3$. The first two of these are carried out in appendix~\ref{app:threeloop}. With $s=\sqrt z$ and $\chi_3(s)=\frac12[\mathrm{Li}_3(s)-\mathrm{Li}_3(-s)]$, the function $\calI_3$ has the closed form
\begin{gather}
\calI_3(z)=\frac1s\Big[\chi_3(s)-\ln s\,\chi_2(s)+\tfrac12\ln^2s\,\artanh s-\frac{\pi^2}{12}\artanh s+\tfrac12A(s)\Big],\label{eq:calI3}\\
A(s)=\int_0^s\!\frac{dx}{1-x^2}\Big[2x\chi_2(x)-2x\ln x\,\artanh x-\ln x\,\ln(1-x^2)-\tfrac12\Li(x^2)\Big],\label{eq:Adef}
\end{gather}
where $A=H_{-1,1,0}(s)-H_{1,-1,0}(s)$ is a combination of two harmonic polylogarithms of weight three \cite{RV2000}. With $\calO$ acting on $z\calI_3$, the three-loop coefficient is
\begin{equation}\label{eq:C3}
C_3(z)=-\frac{1}{64\pi^2z}\Big[2(3-z)\calI_3(z)+4z\calI(z)-\tfrac12\Li(z)-\tfrac12\ln z\ln(1-z)+\tfrac14\ln^2z+\ln z-\frac{\pi^2}{6}\Big],
\end{equation}
with the structure of \eqref{eq:KKclosed}: the function of highest weight, now $\calI_3$ in place of $\calI$, comes with the factor $3-z$, and the derivatives in $\calO$ acting on $\calI_3$, each of which lowers the weight by one, produce the terms of lower weight.

As $z\to0$, $C_3/C_2\to\frac14\ln(1/z)+\frac14$, where $C_2$ is the coefficient of $a^2\ln\Lambda$ in \eqref{eq:KKclosed}. Since the term of order $a^3$ is $(C_3/C_2)\,a\ln\Lambda$ times the term of order $a^2$, at small angles their ratio grows as the double logarithm $\frac14a\ln\Lambda\ln(1/z)$. There the fixed-order result \eqref{eq:KKclosed} holds only while $a\ln\Lambda\ln(1/z)\ll1$, a stronger condition than $a\ln\Lambda\ll1$. The ratio $C_3/C_2$ is positive for $0<z\leq1$, as we prove in appendix~\ref{app:threeloop}, so the term of order $a^3$ has the same sign as the term of order $a^2$ and adds to the growth.

\paragraph{Large windows.} For $a\ln\Lambda\gg1$ the factor $\Lambda^{j(\nu)-1}$ in \eqref{eq:LL} confines the $\nu$ integral to small $\nu$. Near $\nu=0$, where the spin is largest, $j=1+4a\ln2-\frac72\zeta(3)\,a\nu^2+O(\nu^4)$, so this factor is $\Lambda^{4a\ln2}$ times a Gaussian in $\nu$ of width of order $(a\ln\Lambda)^{-1/2}$, and the integral can be done by the saddle-point method. Since the weight vanishes like $\nu^2$ at $\nu=0$, the integral gives the growth $\Lambda^{4a\ln2}(a\ln\Lambda)^{-3/2}$, with the power $-\frac32$ of $a\ln\Lambda$ rather than the $-\frac12$ of a weight that does not vanish. At small angles the conical function $P_{-\frac12+\frac{i\nu}{2}}(2/z-1)$ oscillates in $\nu$, with a phase that grows as $\frac12\nu\ln(1/z)$, and the saddle-point form holds only if this phase changes little across the Gaussian, that is, if $a\ln\Lambda\gg\ln^2(1/z)$. At two loops the Regge term has spin exactly 1 and gives a logarithm of the window. Once the leading logarithms are resummed, the spin at $\nu=0$ is the planar intercept $j(0)=1+4a\ln2$, and the logarithm becomes a power of the window.

\paragraph{Beyond weak coupling.} The growth of $\vev{K_1K_2}$ as a power of the window is not special to weak coupling. The planar intercept $j(0)$ is $1+4a\ln2$ at weak coupling and $2-1/(\pi\sqrt a)$ at strong coupling \cite{BPST2007,CGP2012}. Between these limits it has been computed from the quantum spectral curve up to $a\approx25$, that is $g^2N\approx10^3$, and it is above 1 at every coupling computed \cite{GLS2015}. The detector kernels that turn the tail of the Mellin amplitude into the dependence on the window do not depend on the coupling. Each partial wave therefore grows as $\Lambda^{j(\nu)-1}$ at any coupling, where $j(\nu)$ is now the spin of the leading Regge trajectory rather than the BFKL spin, but the coefficient is no longer that of \eqref{eq:LL}. Hence, as long as the Regge amplitude stays far below its black-disk value, so that a single Regge exchange describes it, the planar correlator grows with the time window as $\Lambda^{j(0)-1}$, up to powers of $\ln\Lambda$, all the way from weak to strong coupling.

\section{Discussion}
\label{sec:discussion}

We have computed the logarithm of the window in $\vev{K_1K_2}$ at two loops and identified it in $\Nfour$~SYM with the two-gluon Regge term. The logarithm remains on normalizable wave packets and also implies that the states created by the source lie outside the domain of the smeared time-weighted detector, unless the smearing is uniform. In planar $\Nfour$~SYM, the Regge origin of the logarithm and the BFKL spin determine the leading logarithms beyond two loops, which resum to a power of the window. Two steps, however, are done only in part: we treat the detector's pair terms at tree level without evaluating the other pieces of the soft sector at order $a^2$ one by one, and we derive the exchange of the large-$\tau$ limit with the $t$ integral at the level of exponent counting, without the constants in its bounds. Several cases also lie outside the computation. In a theory that is not conformal, we have not excluded a logarithm in $\vev{E_1K_2}_{\rm sym}$, which on a wave packet would add to the coefficient a term that depends on the position of the packet. In QCD the coupling runs inside the integral $\int d\lambda/\lambda$, and the hadron masses cut this integral off at low energies. Lastly, for a source with several hard partons at lowest order the soft current is a dipole sum, and we expect the same logarithm dipole by dipole.

Several questions about the logarithm itself remain open. Two would test the present results: a check of the spin-1 identification beyond $\Nfour$~SYM, which needs the kernel of stress-tensor detectors in a general conformal theory and, at order $g^4$, the Regge limit of the correlator of two stress tensors in the state created by the source, and a direct computation of the three-loop coefficient from soft partons, which would give the analogue at order $a^3$ of the two-loop equality with the Regge term. A third concerns finite $N$. The planar leading logarithms resum to a power of the window, but at fixed $N$ the Regge amplitude is $1/c_T$ times this power, so it is small at moderate windows and grows until unitarity bounds it. The planar intercept exceeds the bound $J_0\leq1$ that holds at finite $N$ \cite{KKSZ2020a}, so the planar series cannot show whether a residue at $J=1$ survives at finite $N$. An intercept $J_0=1$, expected in a gauge theory \cite{KKSZ2020a}, is the marginal case of the convergence condition and does not by itself decide the question. If the residue at $J=1$ is nonzero, the states created by the source stay outside the domain of $K$. If it vanishes, these states may lie in the domain, and we have not shown which of the two cases holds.

Detectors related to $K$ raise similar questions. Higher moments in time, $\int du\,u^{n+1}T_{uu}$, do not in general have well-defined matrix elements \cite{Besken2021}. Detectors with other powers of the energy have infrared divergences that are explained and renormalized in \cite{CaronHuot2023}, but they are translation invariant, while $K$ is not. The number detector has the same celestial weight as $K$ but is a different operator \cite{HuPasterski2023}. In gravity, which detectors of higher memories exist as quantum operators, and whether they need renormalization, is asked in \cite{HMOPS2026}. Our results answer the analogous question for $K$ in a gauge theory: it is a self-adjoint operator on Fock space, but, unless the smearing is uniform, the states created by the source lie outside its domain, at least in perturbation theory.

It is suggested in \cite{HP2026} that the light-ray operators constructed there may provide new infrared-finite collider observables, particularly in QCD. Among them, the local dilatation $\calD=\int du'\,u'\,T_{uu}^{(2)}$ of \cite{HP2025,HP2026} is $K$ built on the flat section of null infinity instead of the round one (appendix~\ref{app:HP}). A single such dilatation has a finite one-point function in a conformal theory, like those computed in \cite{HP2026}, but the product of two in a physical state diverges at two loops like the logarithm of the window. The results in this paper therefore limit that suggestion but do not refute it: the divergence comes from expanding the two detectors around zero frequency, and at nonzero frequency the correlators are infrared finite \cite{KSZ2022}.

A test of the algebra on $\vev{O\,T\,T\,O}$ in $\Nfour$~SYM is also proposed in \cite{HP2026}. One of the commutators it involves, $[E(n),K(n')]$, is between the average null energy operator and the local dilatation, and the convergence condition of \cite{KKSZ2020a} for $\vev{EK}$ fails in holographic theories at large $N$ and infinite gap, where the Regge intercept is $J_0=2$ \cite{Belin2021}. The test concerns the commutators of these operators, whereas the logarithm found here is in their product, $\vev{K_1K_2}$. At a common window the logarithm of $\vev{K_1K_2}$ at separated directions is real and symmetric, so it cancels in the commutator. The commutators at order $a^2$ are therefore blind to the logarithm: the test can neither detect it nor be spoiled by it. On Fock space at zero frequency, $[K[f],K[g]]=0$ identically, so the test can succeed even though the states created by the source lie outside the domain of $K[f]$ for nonconstant $f$. Since $\vev{K_1K_2}$ fails to converge absolutely, the test needs a prescription with a common window for both detectors. What the test can show about these operators on physical states needs further investigation, for instance a computation of the contact term of the commutator at order $a^2$ at a common window.

\section*{Acknowledgments}
The author formulated the questions, the approach and the main ideas. Under the author's direction, Claude Opus 5.5 and Claude Fable 5.2 (Anthropic) assisted at every stage, including the derivations, the literature search and the drafting; GPT-6.2 Astra (OpenAI) assisted in checking and reviewing the derivations. The author rederived the analytic results and checked the arguments, and is solely responsible for the content.

\appendix
\section{Window-shape independence}
\label{app:kernels}

Here we show that the coefficient of the logarithm does not depend on the shape of the window, for gluons, Weyl fermions and scalars with any improvement term. It suffices to treat the kernels to the first order at finite frequency. On the soft pair let $\mathcal{E}(\omega_1,\omega_2)$ be the correlator of $E_{\omega_1}(n_1)$ and $E_{\omega_2}(n_2)$, up to the overall factor of \eqref{eq:KKsum}. The correlator of two windowed detectors $K_\chi$ is then, up to the same factor,
\begin{equation}\label{eq:windowedKK}
\vev{K_\chi(n_1)K_\chi(n_2)}=\int\!\frac{d\omega_1d\omega_2}{(2\pi)^2}\,\hat W(-\omega_1)\hat W(-\omega_2)\,\mathcal{E}(\omega_1,\omega_2),
\end{equation}
with $\hat W(\omega)=\int du\,e^{i\omega u}W(u)$ and $W(u)=u\chi(u/T)$ as their weight. Since the integral $\int d\omega\,\hat W(-\omega)=2\pi W(0)$ vanishes, a term of $\mathcal{E}$ that does not depend on one of the two frequencies integrates to zero, so \eqref{eq:windowedKK} depends on $\mathcal{E}$ only through its mixed second difference $\mathcal{E}(\omega_1,\omega_2)-\mathcal{E}(\omega_1,0)-\mathcal{E}(0,\omega_2)+\mathcal{E}(0,0)$.

Each kernel of section~\ref{sec:setup}, divided by $\sqrt{PP'}$, is $1+O((P'-P)^2/P^2)$, so at energies much larger than the frequencies, the term of $\mathcal{E}$ in $\omega_1\omega_2$ is the same for all species: its integrand is $-s^*\partial_y^2s/(E_1E_2)$ in the measure $dE_1dE_2$. This term gives $\omega_1\omega_2\ln(M/|\omega|)\int dy\,|\partial_ys|^2$, with $|\omega|$ the common scale of the two frequencies. The remaining part of the integrand falls like the third inverse power of the energies or faster. Its integral therefore converges at large energy and comes from energies of order $|\omega|$; with no other scale, it is homogeneous of degree two in the frequencies, so it has no logarithm. In the region where only one of the two energies is of order $|\omega|$ and the other is much larger, $|y|$ is large, and the integrand falls like $\partial_ys$, which decreases exponentially in $|y|$; the integral over $y$ then converges, and this region adds no logarithm. For gluons, $s$ tends to the constants $s_\pm$ as $y\to\pm\infty$. In these limits $F=s_\pm/(E_1E_2)$ is a product of two eikonal factors, as in independent emission, so the part of $\mathcal{E}$ from these limits has the form $\mathcal{E}_1(\omega_1)\mathcal{E}_2(\omega_2)$. Its mixed second difference is $[\mathcal{E}_1(\omega_1)-\mathcal{E}_1(0)]\,[\mathcal{E}_2(\omega_2)-\mathcal{E}_2(0)]$, and neither factor has a logarithm.

So the coefficient of $\ln(M/|\omega|)$ in $-\partial_{\omega_1}\partial_{\omega_2}\mathcal{E}$ is $-\int dy\,|\partial_ys|^2$. With $\hat W(\omega)=T^2w(\omega T)$ and $\int d\omega\,\hat W(-\omega)\,\omega=-2\pi i\chi(0)$, the windowed correlator \eqref{eq:windowedKK} has the same coefficient of $\ln(MT)$, for every species and every window with $\chi(0)=1$. The shape of the window and the higher orders of the kernels therefore change only the constant.

\paragraph{The flux in time.} The time-domain amplitude \eqref{eq:timeamp} and the integral \eqref{eq:Tnorm} over the times rest on the energy flux of a single parton as a function of the retarded time. $K$ acts as $-i(P\partial_P+\tfrac12)$ on the wavefunction normalized to unit norm, and as $-i(P\partial_P+1)$ on a relativistic amplitude, which is the normalized wavefunction divided by $\sqrt{P}$, the factor from the measure $P\,dP$. With a source at the origin in the ket and a momentum eigenstate in the bra it acts on the ket amplitude alone. Since the kernel $\sqrt{PP'}$ factorizes into $\sqrt P\sqrt{P'}$, the flux in time is the modulus squared of a single Fourier integral: with $\hat F$ the normalized wavefunction and $\mathcal T(u)=\int dP\,\sqrt P\,\hat F(P)\,e^{-iPu}$, $E(u)=|\mathcal T(u)|^2/2\pi$. Then $\int du\,E(u)=\vev{P}$, and
\begin{equation}\label{eq:firsttime}
\int\!du\,u\,E(u)=\int\!dP\,\hat F^*(-i)\big(P\partial_P+\tfrac12\big)\hat F .
\end{equation}
On two partons at separated directions this gives \eqref{eq:timeamp}, where the normalized wavefunction is $\hat F=s(y)/\sqrt{E_1E_2}$, so that $E_1E_2F=s(y)$.

\section{Reduction of the angular averages}
\label{app:reduction}

Here we derive the closed form \eqref{eq:Gclosed} of $\vev{G}$ and, by the same steps, the averages \eqref{eq:Uaverages} for charge flux. From \eqref{eq:species}, the average of $G$ is $\vev{G}=n_s/6+n_f/12+1/3-n_s\calA_1/2-n_f\calA_2/8-\calA_3$, with $\calA_1=\vev{h_1}$, $\calA_2=\vev{\cos\phi\,h_2}$ and $\calA_3=\vev{\cos2\phi\,h_1}$. Let $v^2=1-z(1-w^2)$ and $w=(E_1-E_2)/(E_1+E_2)$. We average over the hard axis at fixed $y$ first, in an orthonormal frame adapted to the combined direction $E_1\vec n_1+E_2\vec n_2$. The remaining rational integrals give
\begin{equation}\label{eq:reduction}
\calA_i=\int_{-1}^1\!dw\,\Big[\alpha_i\frac{\artanh v}{v}+\beta_i\Big],
\end{equation}
with $\alpha_i$ and $\beta_i$ rational in $w^2$ and $z$, whose denominators are powers of $v^2$ up to $v^8$. To integrate them, we write them as Laurent polynomials in $v^2$ and use the recursion, down to $m=-4$, for $\mathcal K_m=\int_0^1dw\,v^{2m-1}\artanh v$ that follows from integrating $\partial_w[wv^{2m-1}\artanh v]$ by parts:
\begin{equation}\label{eq:recursion}
2m\,\mathcal K_m-(2m-1)(1-z)\,\mathcal K_{m-1}=-\tfrac12\ln z-\int_0^1\!dw\,\Big[\frac{v^{2m-2}-1}{1-w^2}-v^{2m-2}\Big],\qquad\mathcal K_0=\calI(z).
\end{equation}
The logarithm comes from the endpoint $w=1$, where $\artanh v$ diverges. Terms in $\arcsin\sqrt z$ appear in intermediate steps at $m<0$ and cancel. The recursion gives
\begin{equation}\label{eq:Aclosed}
\begin{aligned}
\calA_1&=\frac{(1+z)(3z^2-14z+3)}{64z}\calI+\frac{3z^2+32z-3}{64z}+\frac{(z-3)(3z-1)}{128z}\ln z,\\
\calA_2&=\frac{5z^3-21z^2+3z-3}{16z}\calI+\frac{(1+z)(5z+3)}{16z}+\frac{(1+z)(5z-3)}{32z}\ln z,\\
\calA_3&=\frac{35z^3-75z^2+21z+3}{64z}\calI+\frac{35z^2-3}{64z}+\frac{(1+5z)(3+7z)}{128z}\ln z,
\end{aligned}
\end{equation}
and inserted in $\vev{G}$ they give \eqref{eq:Gclosed}. The coefficient of $1/z$ in the term with $\calI$ is proportional to $n_s-n_f+2$. At small $z$ it cancels against the other two terms for any content, since $\calI=\sum_kz^k[(2k+1)^{-2}-\ln z/(4k+2)]$. In $\Nfour$~SYM the integrands also sum to \eqref{eq:KKw}. For charge flux the same steps give \eqref{eq:Uaverages} from
\begin{equation}
\vev{U_s}=\int_{-1}^1\!dw\,\Big[\frac{z(w^2-1)(w^2z^2+w^2z-z^2+3z-2)}{2v^4}\,\frac{\artanh v}{v}-\frac{z(w^2z-3w^2-z+1)}{2v^4}\Big],
\end{equation}
and $\vev{U_f}$ follows from it as $4z\calI(z)-\vev{U_s}$.

\section{The kernels at small frequency}
\label{app:identity}

\paragraph{The identity for $\vev{EK}$.} The kernels of \cite{KSZ2022} expanded at small frequency have the coefficients $c_{00}=1/z^2$ and $c_{01}=(1-j_1)/z^2$ in the representation \eqref{eq:kernel}. Since $\calM$ is even in $\tau$ and $j_1=t/2+i\tau$, only the part of $c_{01}$ even in $\tau$ survives, namely $c_{01}\to c_{00}(1-t/2)$, with no $\tau^2$ term, unlike $c_{11}$. With $g(X)=z^2(1-z)\mathrm{EEC}(z)$, which is the integral of $(\pi/2\sin\pi t)X^tt^2\calM$, the factor $t$ acts as $X\partial_X$ on $g$ alone. In this paragraph and the next, $w_2$ denotes the second argument of EEC, which is twice the frequency of the detector. This gives
\begin{equation}\label{eq:dEEC}
\partial_{w_2}\mathrm{EEC}(0,w_2)\big|_{w_2=0}=\frac{(1-\tfrac12X\partial_X)\,g(X)}{z^2(1-z)},
\end{equation}
and with $\vev{E_1K_2}=-(i/4\pi^2)\partial_{w_2}\mathrm{EEC}$ it is \eqref{eq:EKidentity}. This holds when the odd part $\tau\calM$ is absolutely summable, which is when $J_0<2$.

\paragraph{Control at $J_0=2$.} At $J_0=2$ the odd part is marginal and the two one-sided derivatives at $w_2=0$ can differ. At $\omega_1=0$ the kernel of \cite{KSZ2022} gives the strong-coupling correlator in closed form, $\mathrm{EEC}=(1+w_2)^2(w_2^2+6w_2+6)/[12(1+zw_2)^3]$ for $w_2<0$ and $(w_2^2+6w_2+6)/12$ for $w_2>0$, so $\partial_{w_2}\mathrm{EEC}$ is $\tfrac32(1-z)$ for $w_2\to0^-$ and $\tfrac12$ for $w_2\to0^+$. The right side of \eqref{eq:dEEC}, on $\mathrm{EEC}=\tfrac12$, is $1-\tfrac34z$, the mean of the two. The two one-sided values differ from it by $\pm(\tfrac12-\tfrac34z)$, the integral of the tail, which enters with the sign of $-w_2$. The mean agrees with the right side whenever both one-sided derivatives exist, since the crossing relation of these correlators at $\omega_1=0$ fixes their sum to $(4-3z)\,\mathrm{EEC}+z(1-z)\,\partial_z\mathrm{EEC}$, which is twice the right side.

\paragraph{The kernel at large $\tau$.} At small frequencies and large $\tau$, with $b_i=2\tau\omega_i/M$ fixed, the energy kernel of \cite{KSZ2022} in \eqref{eq:kernel}, normalized to $1/z^2$ at zero frequency, tends to
\begin{equation}\label{eq:Hscaling}
H(b_1,b_2)=e^{i(b_1-b_2)}\dint{s}\,\frac{\Gamma(-s)\Gamma(1-t-s)}{\Gamma(1-t)}\,\xi^s\,\frac{h(s;b_1,b_2)}{6t^2z^2(1-t^2)},
\end{equation}
where $\xi=(1-z)b_1b_2$ for $b_1b_2>0$ and $\xi=e^{\mp i\pi}(1-z)|b_1b_2|$ for $\pm b_1>0>\pm b_2$, and
\begin{equation}\label{eq:hpoly}
\begin{aligned}
h&=h^{(0)}+b_1b_2\,h^{(1)}+(1-z)(5-z)\,b_1^2b_2^2-2i(b_1-b_2)\big[s(s-t)(s+t-1)\\
&\quad-(1-z)(s-t+1)\,b_1b_2\big],\\
h^{(0)}&=-s^2(s-1)^2+6t(1-t^2)(s+t),\qquad h^{(1)}=-4s^2+2s(t-1)(z-3).
\end{aligned}
\end{equation}
The poles at $s=n$ give the expansion in the frequencies, whose term in $b_1b_2$ is the large-$\tau$ coefficient of \eqref{eq:kernel}. The poles at $s=1-t+n$ give the powers $(b_1b_2)^{1-t+n}$, which do not appear in any expansion in the frequencies.

\paragraph{The limit from the exact kernel.} Let $w_i=2\omega_i/M$ and $Y=(1-z)w_1w_2/[(1+w_1)(1+w_2)]$. At equal signs the scalar kernel of \cite{KSZ2022}, divided by its value at zero frequency, is
\begin{equation}\label{eq:kernelMB}
(1+w_1)^{-j_2}(1+w_2)^{-j_1}\dint{s}\,\frac{\Gamma(-s)\Gamma(1-t-s)\Gamma(j_1+s)\Gamma(j_2+s)}{\Gamma(1-t)\Gamma(j_1)\Gamma(j_2)}\,Y^s,
\end{equation}
and the energy kernel follows from it by a differential operator of second order in each $w_i$. Every derivative acts only on $(1+w_1)^{-j_2}(1+w_2)^{-j_1}Y^s$, which carries all the dependence on the $w_i$, so the energy kernel is the same integral with a rational function of $s$ inserted. We set $w_i=b_i/\tau$ and let $\tau\to\pm\infty$ at fixed $s$, $b_i$ and $t$. Then the prefactor tends to $e^{i(b_1-b_2)}$. The ratio $\Gamma(j_1+s)\Gamma(j_2+s)/[\Gamma(j_1)\Gamma(j_2)]$ is $\tau^{2s}[1+O(\tau^{-2})]$, so its product with $Y^s$ tends to $\xi^s$, since $\tau^2Y\to(1-z)b_1b_2$. The rational function that the derivatives insert tends to $h/[6t^2z^2(1-t^2)]$, which gives \eqref{eq:Hscaling} at equal signs.

For $w_1>0>w_2$ the kernel has $\Gamma(1-j_1)/\Gamma(1-j_1-s)$ in place of $\Gamma(j_1+s)/\Gamma(j_1)$ and $-Y$ in place of $Y$. By the reflection formula the two ratios differ by the factor $\sin\pi(j_1+s)/\sin\pi j_1$, and at large $|\tau|$ this factor is $e^{-i\pi s}$ for $\tau>0$ and $e^{i\pi s}$ for $\tau<0$, up to terms of order $e^{-2\pi|\tau|}$. The opposite case, $w_1<0<w_2$, follows by exchanging the detectors. The limit holds term by term in the two series of poles, which converge for every $\xi$, and for negative $\xi$ their sum defines the integral in \eqref{eq:Hscaling}.

The first correction to this large-$\tau$ limit is $1/\tau$ times an integral of the same form, with a polynomial of two more powers of the $b_i$ and one more of $t$ in place of $h$, and it is proportional to $b_1+b_2$. So the limit needs $b_i^2\ll|\tau|$ and $|t\,b_i|\ll|\tau|$. At fixed frequencies these are $|\tau|\ll(M/2\omega_i)^2$ and $|t|\ll M/2\omega_i$, the former far beyond the scale $M/2\omega_i$ where the kernel cuts off the $\tau$ integral.

\section{The two-loop Mellin amplitude and its tail}
\label{app:tail}

At two loops the function of \eqref{eq:Mellin} is \cite{Eden2012,BHKSZ2014c}
\begin{equation}\label{eq:Phi2loop}
\Phi\big|_{a^2}=\frac{1+U+V}{32}\,\bar\Phi^2+\frac18\Big[\Phi_2(U,V)+\frac1U\,\Phi_2\Big(\frac VU,\frac1U\Big)+\frac1V\,\Phi_2\Big(\frac1V,\frac UV\Big)\Big],
\end{equation}
with $\bar\Phi$ the box function and $\Phi_2$ the two-loop ladder function, each normalized by its Mellin amplitude below. In the Mellin representation $\int ds_1ds_2/(2\pi i)^2\,(\cdots)\,U^{s_1}V^{s_2}$ and with $G_3(s_1,s_2)=[\Gamma(-s_1)\Gamma(-s_2)\Gamma(1+s_1+s_2)]^2$, the amplitude of $\bar\Phi$ is $G_3$, and that of $\bar\Phi^2$ is the convolution of two such factors, as in \cite{BHKSZ2014c},
\begin{equation}\label{eq:Dconv}
D(A_1,A_2)=\int\!\frac{ds_1ds_2}{(2\pi i)^2}\,G_3(s_1,s_2)\,G_3(A_1-s_1,A_2-s_2),
\end{equation}
on contours where the arguments of all six $\Gamma$ functions have positive real parts; they exist when the real parts of $A_1$, $A_2$ and $-2-A_1-A_2$ are negative. The amplitude of $\Phi_2$ is known in closed form \cite{Allendes2013,BhatMaity2023}: it is $\tfrac12G_3(s_1,s_2)\,B(s_1,s_2)$, with the function $B$ given by
\begin{equation}\label{eq:Bladder}
B(s_1,s_2)=\psi'(-s_1)+\psi'(-s_2)-\big[\psi(-s_1)-\psi(-s_2)\big]^2-\pi^2 .
\end{equation}
In this normalization $\Phi_2$ is 16 times the ladder function of \cite{BHKSZ2014c}, and \eqref{eq:Bladder} reproduces the twofold Mellin--Barnes integral given there for its amplitude: with $\Gamma(-s)/\Gamma(1-s)=-1/s$ and a Feynman parameter, that integral is $G_3(s_1,s_2)\,(1+s_1+s_2)\,S/16$ with $S=\sum_{k,l\geq0}[(k+l+1)(k-s_1)(l-s_2)]^{-1}$, and $(1+s_1+s_2)S=B(s_1,s_2)/2$. The two sides of this identity obey the same recurrence under $s_1\to s_1-1$, so their difference is a periodic entire function of $s_1$; it grows at most like a power, so it is a constant, and both sides vanish on the line $s_1+s_2=-1$. Since the exponents of $U$ and $V$ in \eqref{eq:Mellin} are $-1-t$ and $j_2$, the amplitude at order $a^2$ is
\begin{equation}\label{eq:M2}
\begin{split}
\calM&=a^2\,\frac{D(-1-t,j_2)+D(-2-t,j_2)+D(-1-t,j_2-1)}{32\,[\Gamma(1+t)\Gamma(1-j_1)\Gamma(1-j_2)]^2}\\
&\quad+a^2\,\frac{B(-1-t,j_2)+B(j_2,j_1)+B(j_1,-1-t)}{16\,j_1^2j_2^2},
\end{split}
\end{equation}
with the three convolutions from the terms $1$, $U$ and $V$ of \eqref{eq:Phi2loop}, respectively. On the contour of \eqref{eq:Mellin}, with $-1<\mathrm{Re}\,t<0$, each lies inside the domain of \eqref{eq:Dconv}, and $\calM$ is even in $\tau$.

\paragraph{The tail.} Let $j_2=t/2-i\tau$ with $\tau\to\infty$ at fixed $t$. The convolution \eqref{eq:Dconv} splits exactly into two integrals that can be expanded at large $\tau$: one gives the tail and the other the logarithm in its remainder, $\calM-m(t)|\tau|^{-3}$. The ladder terms are explicit and come last. Let $D_2$ and $D_1$ be the same convolution with the extra factors $(1-e^{2\pi is_2})(1-e^{2\pi i(A_2-s_2)})$ and $-2(1-e^{2\pi i(A_2-s_2)})$ in the integrand, respectively. Without them the integrand is symmetric under $(s_1,s_2)\to(A_1-s_1,A_2-s_2)$, so
\begin{equation}\label{eq:Dsplit}
\big(e^{2\pi iA_2}-1\big)\,D(A_1,A_2)=D_2+D_1 .
\end{equation}
In position space this is the rotation of the $V$ contour of the Mellin transform by $2\pi$: the factor $e^{2\pi is_2}-1$ is the monodromy around $V=0$, and $D_2$ and $D_1$ are the transforms of $(\delta\bar\Phi)^2$ and of $2\bar\Phi\,\delta\bar\Phi$. The monodromy of the box is exact, $(\delta\bar\Phi)^2=-4\pi^2\beta(w_R)^2/U$ with $w_R=(1+U-V)/\sqrt U$, not only in the Regge limit; in Mellin space this is Barnes's first lemma. Here $\beta$ is the function of \eqref{eq:f2}, with the representation
\begin{equation}\label{eq:betaMB}
\beta(w_R)=\dint{s}\,\Gamma(-s)^2\,\Gamma(1+2s)\,w_R^{-1-2s},\qquad-\tfrac12<\mathrm{Re}\,s<0 .
\end{equation}
Squaring it with the Mellin convolution theorem gives $\beta(w_R)^2=\dint{s}\,\Gamma(-2s)\,\mathcal J(-1-s)\,w_R^{2s}$ for $-1<\mathrm{Re}\,s<0$, and a beta integral over $U$ and $V$ then gives
\begin{equation}\label{eq:D2}
\begin{split}
D_2&=-4\pi^2e^{i\pi A_2}\,\Gamma(-A_2)\dint{s}\,\mathcal J(-1-s)\,\Gamma(-1-A_1-s)\,\Gamma(1+A_1+A_2-s),\\
\mathcal J(A)&=\frac{1}{\Gamma(2A+2)}\dint{s}\,\Gamma(-s)^2\,\Gamma(s-A)^2\,\Gamma(2s+1)\,\Gamma(2A-2s+1),
\end{split}
\end{equation}
on contours with
\begin{gather}
-1<\mathrm{Re}\,s<\min(0,\,-1-\mathrm{Re}\,A_1,\,1+\mathrm{Re}\,A_1+\mathrm{Re}\,A_2),\label{eq:contourD2}\\
\max(-\tfrac12,\,\mathrm{Re}\,A)<\mathrm{Re}\,s<\min(0,\,\mathrm{Re}\,A+\tfrac12),\label{eq:contourJ}
\end{gather}
respectively; the second exists for $-1<\mathrm{Re}\,A<0$.

In $D_2$ the variable $\tau$ enters through $e^{i\pi A_2}\Gamma(-A_2)\Gamma(1+A_1+A_2-s)$ only, and the integrand falls like $\tau^{-s}$ as the contour is moved to the right. Since $\beta$ is analytic at $w_R=0$, $\mathcal J(-1-s)$ is analytic for $\mathrm{Re}\,s>-1$, so, apart from poles of $\Gamma(1+A_1+A_2-s)$ at $\mathrm{Im}\,s\simeq-\tau$ with exponentially small residues, the contour meets only the simple poles of $\Gamma(-1-A_1-s)$, each smaller than the one before by $1/\tau$. The first of these poles, at $s=-1-A_1$, gives
\begin{equation}\label{eq:Dtail}
D(A_1,j_2)=8\pi^3e^{-2\pi ij_2}e^{-i\pi A_1}\,\tau^{2A_1+1}\mathcal J(A_1)\big[1+O(\tau^{-1}\ln\tau)\big].
\end{equation}

The logarithm in the remainder comes from $D_1$. Relative to the $\Gamma$ functions in the denominator of \eqref{eq:M2}, its extra factor leaves the integrand unsuppressed only where $\mathrm{Im}\,s_1$ and $\mathrm{Im}\,s_2$ are of order one, so that one box carries all of $\tau$. There it is of relative order $\tau^{-1-2s_1}$, and the first singularity to the right of the $s_1$ contour is the double pole of $\Gamma(-s_1)^2$ at $s_1=0$, so $D_1/D_2=-2\Gamma(-A_1)^2\ln\tau/[\pi\mathcal J(A_1)\tau]$ at leading order.

The terms $D(-1-t,j_2)$ and $D(-1-t,j_2-1)$ have the same leading form, $D(-2-t,j_2)$ is smaller by $\tau^{-2}$ because its first pole is at $s=1+t$, and
\begin{equation}\label{eq:Gammatail}
[\Gamma(1-j_1)\Gamma(1-j_2)]^2\simeq4\pi^2\,\tau^{2-2t}\,e^{-2\pi ij_2}\,e^{i\pi t}.
\end{equation}
The ladder terms are explicit: with $\psi(-j_{1,2})=\ln\tau\mp i\pi/2+O(\tau^{-1})$, $B(j_2,j_1)=O(\tau^{-1})$ and the other two give $-2\ln^2\tau$ at leading order. So, for $-1<\mathrm{Re}\,t<0$,
\begin{equation}\label{eq:Mtail}
\calM=-\frac\pi8\,a^2\,\frac{\mathcal J(-1-t)}{\Gamma(1+t)^2}\,|\tau|^{-3}-\frac{a^2}{8}\,|\tau|^{-4}\ln^2|\tau|+O\big(|\tau|^{-4}\ln|\tau|\big).
\end{equation}

The coefficient of $|\tau|^{-3}$ in \eqref{eq:Mtail} is $m(t)$, written as the Mellin transform of $\mathcal{R}(w_R)$ in \eqref{eq:mclosed}: the square of \eqref{eq:betaMB} gives the transform $\int_0^\infty(dw_R/w_R)\,w_R^{-2t}\beta^2$ as $\tfrac12\Gamma(-2t)\mathcal J(-1-t)$. The tail of the amplitude and the residue fixed by the Regge limit are the same integral, and since this integral comes from a simple pole there is no logarithm at order $|\tau|^{-3}$. The transform of $\beta^2$ also gives \eqref{eq:ftransform} without an exchange of limits: its right side, with this coefficient, is the inverse Mellin transform of $-(\pi^2/4)a^2\,\Gamma(-2t)\mathcal J(-1-t)$, which is \eqref{eq:f2}, the position-space limit given by the monodromies.

The ladder terms cannot reach spin 1 in position space either. With $U=\sigma^2$ as in section~\ref{sec:regge}, each is a pure polylogarithm of weight four, in one of the three channels, over the common factor $\sqrt{(1+U-V)^2-4U}$, which is $2\sigma\sinh\rho$ with $2\cosh\rho=(1+U-V)/\sigma$, and their monodromies are elementary. Around $V=0$ the term $\Phi_2(1/V,U/V)/V$ changes by exactly $-i\pi\beta\,(\ln^2\sigma-\rho^2)/\sigma$, and the term $\Phi_2(U,V)$ changes by $-i\pi\beta\ln^2\sigma/\sigma$ up to terms of order $\ln\sigma/\sigma$, while the change of $\Phi_2(V/U,1/U)/U$ is only of order $\ln\sigma$. On the second sheet the ladder terms therefore grow like $\ln^2\sigma/\sigma$ against $\sigma^{-2}$ for $\bar\Phi^2$: at leading order they add $-(i\pi/4)\,a^2\beta\,\sigma\ln^2\sigma$ to $U\Phi$, which vanishes in the Regge limit and is the image of the term of order $|\tau|^{-4}\ln^2|\tau|$ in \eqref{eq:Mtail}.

The exchange of the large-$\tau$ limit with the $t$ integral in \eqref{eq:kernel} rests on two bounds, which we establish at the level of exponent counting. By Stirling's formula and the triangle inequality, the integrands of $D_2$ and $D_1$, divided by $e^{2\pi iA_2}-1$, are bounded in modulus by the $\Gamma$ functions in the denominator of \eqref{eq:M2} times a power of the imaginary parts, uniformly on the contours, and the ladder part of the amplitude is explicit. So $|\calM|$ is bounded by a power of $1+|\mathrm{Im}\,t|+|\tau|$ on the contour of \eqref{eq:Mellin}, and for $|\tau|\geq2(1+|\mathrm{Im}\,t|)$ the remainder is bounded by a power of $1+|\mathrm{Im}\,t|$ times $|\tau|^{-4}\ln^2|\tau|$. The weight of the $t$ integral falls like $e^{-\pi|\mathrm{Im}\,t|}$, so by dominated convergence the integral over $|\tau|<\Lambda$ is $\ln\Lambda/\pi$ times the $t$ integral of the coefficient of the tail, plus a constant, up to terms that vanish as $\Lambda\to\infty$: it has the logarithm of $\Lambda$ with the coefficient of \eqref{eq:KKclosed} and no $\ln^2\Lambda$. The second bound holds only for $|\tau|\geq2(1+|\mathrm{Im}\,t|)$. The rest of the contour contains the lines $\mathrm{Im}\,t=\pm2\tau$, on which the imaginary part of $j_1$ or $j_2$ vanishes and the amplitude falls only like $1/|\tau|$. The first bound suffices there, since on this part of the contour the weight of the $t$ integral is at most of order $e^{-\pi|\tau|/2}$, and on the two lines it is of order $e^{-2\pi|\tau|}$.

\section{The leading logarithm at three loops}
\label{app:threeloop}

Here we derive the closed form \eqref{eq:calI3} of $\calI_3$, which determines $C_3$ through \eqref{eq:C3}, and show that $C_3$ is negative at every angle. The $\nu$ integral that gives the residue $m_3(t)$ can be done in closed form. In \eqref{eq:mnu} the ratio of $\Gamma$ functions is
\begin{equation}\label{eq:Gammaphi}
\frac{\Gamma\big(t+\frac{1+i\nu}{2}\big)\Gamma\big(t+\frac{1-i\nu}{2}\big)}{\Gamma(1+t)^2}=2\int_0^{\pi/2}\!d\phi\,\cot\phi\,\sin^{2t}\!\phi\,P_{-\frac12+\frac{i\nu}{2}}\Big(\frac{2}{\sin^2\!\phi}-1\Big).
\end{equation}
We can write the spin and the conical function as
\begin{gather}
j(\nu)-1=a\Big[2\psi(1)-2\psi\Big(\frac{1+i\nu}{2}\Big)+i\pi\tanh\frac{\pi\nu}{2}\Big],\label{eq:jnu}\\
P_{-\frac12+\frac{i\nu}{2}}=\frac{i}{\pi}\coth\frac{\pi\nu}{2}\,\Big[Q_{-\frac12+\frac{i\nu}{2}}-Q_{-\frac12-\frac{i\nu}{2}}\Big],\label{eq:PQ}
\end{gather}
with $Q$ the Legendre function of the second kind. Since the weight $\tanh^2(\pi\nu/2)$ of \eqref{eq:mnu} and the spin $j(\nu)$ are even in $\nu$, the two terms of \eqref{eq:PQ} contribute equally to the $\nu$ integral, so we can work with twice the first term. For $\phi<\pi/2$ the contour then closes on the poles at $\nu=-i,-3i,\dots$, and the residues sum under $Q_{-\frac12+\frac{i\nu}{2}}(x)=\int_0^\infty d\theta\,(x+\sqrt{x^2-1}\cosh\theta)^{-\frac12-\frac{i\nu}{2}}$ to an integral of logarithms over $\theta$, which gives dilogarithms whose derivative in $\phi$ is elementary.

Write the weight as $\tanh^2(\pi\nu/2)=1-1/\cosh^2(\pi\nu/2)$. At fixed $\phi$ the $\nu$ integral converges absolutely for the term with $1/\cosh^2(\pi\nu/2)$, but only conditionally for the term with 1. The conditionally convergent integral may still be done at fixed $\phi$: the representation
\begin{equation}\label{eq:jint}
\frac{j(\nu)-1}{2a}=\int_0^\infty\!dr\,\frac{\cos\nu r-e^{-r}}{\sinh r}
\end{equation}
of the spin and the Mehler--Dirichlet integral for the conical function give
\begin{equation}\label{eq:common}
\int_{-\infty}^\infty\!d\nu\,\frac{j(\nu)-1}{2a}\,P_{-\frac12+\frac{i\nu}{2}}\Big(\frac{2}{\sin^2\!\phi}-1\Big)=2\tan\phi\int_0^\infty\!\frac{dr}{\sinh r}\,\bigg[\frac{\Theta(1-\tan\phi\,\sinh r)}{\sqrt{1-\tan^2\!\phi\,\sinh^2r}}-e^{-r}\bigg],
\end{equation}
with $\Theta$ the step function. The bracketed expression vanishes at $r=0$, and with the factor $\cot\phi\,\sin^{2t}\!\phi$ of \eqref{eq:Gammaphi} the integral over $r$ and $\phi$ converges absolutely in the strip $-\frac12<\mathrm{Re}\,t<0$ of \eqref{eq:mnu}. The result therefore does not depend on the order of the $\nu$ and $\phi$ integrations. At order $a^2$ the same steps, without the factor $j(\nu)-1$, give back the closed form \eqref{eq:mclosed} of $m(t)$.

The tail of the Mellin amplitude at large $|\tau|$, now with its term at order $a^3$, is $\calM\to|\tau|^{-3}[m(t)+a^3m_3(t)\ln|\tau|+O(a^3,\,a^4\ln^2|\tau|)]$, with
\begin{equation}\label{eq:m3}
m_3(t)=\pi\int_0^{\pi/2}\!d\phi\,\sin^{2t}\!\phi\int_0^\phi\!d\phi'\,\big(\ln\sin\phi'+\phi'\tan\phi'\big),
\end{equation}
the analogue of \eqref{eq:mclosed}. We insert $m_3$ in place of $m$ into the transform \eqref{eq:Psi}, do the $t$ integral and integrate by parts. With $s=\sqrt z$ this gives
\begin{equation}\label{eq:calI3int}
s\,\calI_3(z)=-\int_0^{\pi/2}\!d\phi\,\arctan(s\cot\phi)\,\big(\ln\sin\phi+\phi\tan\phi\big),
\end{equation}
and the same steps at order $a^2$ give $s\,\calI(z)=\int_0^{\pi/2}d\phi\,\arctan(s\cot\phi)$. The function $s\calI_3$ has weight three, and its derivative involves only functions of weight two,
\begin{equation}\label{eq:dI3}
(1-s^2)\,\partial_s(s\calI_3)=s^2\calI-\frac14\Li\Big(1-\frac{1}{s^2}\Big)-\frac{\pi^2}{8}.
\end{equation}
Dividing by $1-s^2$ and integrating from $s=0$, where $s\calI_3$ vanishes, raises the weight back to three. The integrand is a sum of functions of weight two times $1/(1-s)$ or $1/(1+s)$. Written in harmonic polylogarithms of $s$, which are iterated integrals with the kernels $1/s$ and $1/(1\mp s)$, it then integrates in their shuffle algebra \cite{RV2000} to the closed form \eqref{eq:calI3}.

To show that $C_3$ has the same sign as $C_2$, we close the contour of \eqref{eq:Psinu} on the same poles, now with the representation
\begin{equation}\label{eq:Qr}
Q_{-\frac12+\frac{i\nu}{2}}\Big(\frac2z-1\Big)=\frac12\int_{\sinh(r/2)>\sqrt X}\frac{dr\,e^{-i\nu r/2}}{\sqrt{\sinh^2(r/2)-X}}.
\end{equation}
By \eqref{eq:Oeigen} the operator $\calO$ acts on $e^{-i\nu r/2}$ as $\partial_r^2+\frac34$, and with $h=\tanh(r/2)$ the result is
\begin{equation}\label{eq:OPsipos}
\begin{gathered}
\calO\Psi=\frac18\int_{\sinh(r/2)>\sqrt X}\frac{dr}{\cosh(r/2)}\,\frac{a^2d_2(r)+a^3d_3(r)\ln\Lambda}{\sqrt{\sinh^2(r/2)-X}},\\
d_2=\frac r2(1+h^2)-h,\qquad d_3=\frac{1+h^2}{2}\int_0^r\!dr'\,r'\tanh\frac{r'}{2}+h+\frac r2(1-h^2)-rh^2,
\end{gathered}
\end{equation}
up to $O(a^3,\,a^4\ln^2\Lambda)$. Here $d_2>0$ since $h<r/2$. Since $\tanh$ is concave, the integral in $d_3$ is at least $r^2h/3$, and then the inequality of the arithmetic and geometric means gives $d_3\geq r\,[2h\sqrt{(1+h^2)/6}-\frac12(3h^2-1)]$, which is positive for $0<h<1$. Therefore $\calO\Psi>0$ at both orders $a^2$ and $a^3$, and $C_2$ and $C_3$ are negative.

\section{Relation to the local dilatation \texorpdfstring{$\calD$}{D}}
\label{app:HP}

The algebras of \cite{HP2025,HP2026} are built from null integrals of the stress tensor and conserved currents in generic conformal field theories in four dimensions. The generators of scaling dimension zero include the local dilatation $\calD$, which is the detector $K$ built on the flat section of null infinity instead of the round one. Here we find how the two are related. Flat retarded coordinates $(u',r',\zeta,\bar\zeta)$ are used in \cite{HP2025,HP2026}, where $\zeta$ is written as $z$; the primes distinguish them from the $u$ and $r$ of section~\ref{sec:setup}. In these coordinates a point in Minkowski space is
\begin{equation}\label{eq:HPcoords}
x^\mu=\tfrac12\big(u'\,\bar n^\mu+r'\,\hat q^\mu\big),\qquad \hat q^\mu=\big(1+|\zeta|^2,\ \zeta+\bar\zeta,\ -i(\zeta-\bar\zeta),\ 1-|\zeta|^2\big),\qquad \bar n^\mu=(1,0,0,-1),
\end{equation}
and the metric is $-du'\,dr'+r'^2d\zeta\,d\bar\zeta$. The vector $\hat q$ is $(1+|\zeta|^2)$ times the null vector $n$, so $\zeta$ is a stereographic coordinate on the sphere, with $d^2n=4\,d^2\zeta/(1+|\zeta|^2)^2$ and $d^2\zeta=d\,\mathrm{Re}\,\zeta\,d\,\mathrm{Im}\,\zeta$. At large radius, \eqref{eq:HPcoords} gives
\begin{equation}\label{eq:HPur}
u'=(1+|\zeta|^2)\,u,\qquad r'=\frac{2r}{1+|\zeta|^2},
\end{equation}
so the sections of constant $r'$ are flat and those of constant $r$ are round. Near null infinity the stress tensor of the outgoing radiation is $T_{\mu\nu}=T_{uu}\,n_\mu n_\nu$. Since $\bar n\cdot n=-2/(1+|\zeta|^2)$, its component along the flat retarded time is
\begin{equation}\label{eq:HPTuu}
T_{u'u'}=\tfrac14\,(\bar n\cdot n)^2\,T_{uu}=\frac{T_{uu}}{(1+|\zeta|^2)^2}.
\end{equation}
The operator $T_{uu}^{(2)}$ of \cite{HP2025,HP2026} is the coefficient of $r'^{-2}$ in this component. With \eqref{eq:HPur} it is $4E(u,n)/(1+|\zeta|^2)^4$, and the generator $\calD$ becomes
\begin{equation}\label{eq:HPD}
\calD(\zeta)=\int\!du'\,u'\,T_{uu}^{(2)}=\frac{4K(n)}{(1+|\zeta|^2)^2}.
\end{equation}
In the same way the average null energy operator in the flat coordinates, $\int du'\,T_{uu}^{(2)}$, is $4E(n)/(1+|\zeta|^2)^3$. With the measure above, $K[f]=\int d^2\zeta\,f\,\calD$ for any function $f$ on the sphere, and $\int d^2\zeta\,\calD$ is the dilatation.

For a state at rest in this frame, \eqref{eq:HPD} gives
\begin{equation}\label{eq:HPDD}
\vev{\calD(\zeta_1)\calD(\zeta_2)}_{\ln}=\frac{16\,\vev{K_1K_2}_{\ln}}{\big[(1+|\zeta_1|^2)(1+|\zeta_2|^2)\big]^2},\qquad z=\frac{|\zeta_1-\zeta_2|^2}{(1+|\zeta_1|^2)(1+|\zeta_2|^2)},
\end{equation}
at separated directions, with a window of width $T$ in $u'$ for $\calD$ and in $u$ for $K$. A window of width $T$ in $u'$ is the same window with width $T/(1+|\zeta|^2)$ in $u$, so \eqref{eq:HPD} holds for the windowed operators when the widths are related in this way. Changing the width in $u$ from $T/(1+|\zeta|^2)$ to $T$ changes the constant under the logarithm and leaves its coefficient as it is. So $\vev{\calD(\zeta_1)\calD(\zeta_2)}$ has the logarithm of the window found in $\vev{K_1K_2}$.

\bibliographystyle{unsrturl-compact}
\bibliography{refs}

\begin{thebibliography}{10}

\bibitem{SveshnikovTkachov1996}
N.~A. Sveshnikov and F.~V. Tkachov.
\newblock {Jets and quantum field theory}.
\newblock \href {https://doi.org/10.1016/0370-2693(96)00558-8} {{\em Phys.
  Lett. B}}, 382:403--408, 1996.
\newblock \href {http://arxiv.org/abs/hep-ph/9512370}
  {\path{arXiv:hep-ph/9512370}}.

\bibitem{Tkachov1997}
Fyodor~V. Tkachov.
\newblock {Measuring multi - jet structure of hadronic energy flow or What is a
  jet?}
\newblock \href {https://doi.org/10.1142/S0217751X97002899} {{\em Int. J. Mod.
  Phys. A}}, 12:5411--5529, 1997.
\newblock \href {http://arxiv.org/abs/hep-ph/9601308}
  {\path{arXiv:hep-ph/9601308}}.

\bibitem{Komiske2018}
Patrick~T. Komiske, Eric~M. Metodiev, and Jesse Thaler.
\newblock {Energy flow polynomials: A complete linear basis for jet
  substructure}.
\newblock \href {https://doi.org/10.1007/JHEP04(2018)013} {{\em JHEP}}, 04:013,
  2018.
\newblock \href {http://arxiv.org/abs/1712.07124} {\path{arXiv:1712.07124}}.

\bibitem{HofmanMaldacena2008}
Diego~M. Hofman and Juan Maldacena.
\newblock {Conformal collider physics: Energy and charge correlations}.
\newblock \href {https://doi.org/10.1088/1126-6708/2008/05/012} {{\em JHEP}},
  05:012, 2008.
\newblock \href {http://arxiv.org/abs/0803.1467} {\path{arXiv:0803.1467}}.

\bibitem{BHKSZ2014a}
A.~V. Belitsky, S.~Hohenegger, G.~P. Korchemsky, E.~Sokatchev, and
  A.~Zhiboedov.
\newblock {From correlation functions to event shapes}.
\newblock \href {https://doi.org/10.1016/j.nuclphysb.2014.04.020} {{\em Nucl.
  Phys. B}}, 884:305--343, 2014.
\newblock \href {http://arxiv.org/abs/1309.0769} {\path{arXiv:1309.0769}}.

\bibitem{BHKSZ2014b}
A.~V. Belitsky, S.~Hohenegger, G.~P. Korchemsky, E.~Sokatchev, and
  A.~Zhiboedov.
\newblock {Event shapes in $\mathcal{N} = 4$ super-Yang-Mills theory}.
\newblock \href {https://doi.org/10.1016/j.nuclphysb.2014.04.019} {{\em Nucl.
  Phys. B}}, 884:206--256, 2014.
\newblock \href {http://arxiv.org/abs/1309.1424} {\path{arXiv:1309.1424}}.

\bibitem{BHKSZ2014c}
A.~V. Belitsky, S.~Hohenegger, G.~P. Korchemsky, E.~Sokatchev, and
  A.~Zhiboedov.
\newblock {Energy-Energy Correlations in N=4 Supersymmetric Yang-Mills Theory}.
\newblock \href {https://doi.org/10.1103/PhysRevLett.112.071601} {{\em Phys.
  Rev. Lett.}}, 112(7):071601, 2014.
\newblock \href {http://arxiv.org/abs/1311.6800} {\path{arXiv:1311.6800}}.

\bibitem{KKSZ2020a}
Murat Kologlu, Petr Kravchuk, David Simmons-Duffin, and Alexander Zhiboedov.
\newblock {Shocks, Superconvergence, and a Stringy Equivalence Principle}.
\newblock \href {https://doi.org/10.1007/JHEP11(2020)096} {{\em JHEP}}, 11:096,
  2020.
\newblock \href {http://arxiv.org/abs/1904.05905} {\path{arXiv:1904.05905}}.

\bibitem{CordovaShao2018}
Clay C{\'o}rdova and Shu-Heng Shao.
\newblock {Light-ray Operators and the BMS Algebra}.
\newblock \href {https://doi.org/10.1103/PhysRevD.98.125015} {{\em Phys. Rev.
  D}}, 98(12):125015, 2018.
\newblock \href {http://arxiv.org/abs/1810.05706} {\path{arXiv:1810.05706}}.

\bibitem{HP2025}
Elizabeth Himwich and Monica Pate.
\newblock {Light-ray Operators and the ${\rm w}_{1+\infty}$ Algebra}, 2025.
\newblock \href {http://arxiv.org/abs/2512.18973} {\path{arXiv:2512.18973}}.

\bibitem{HP2026}
Elizabeth Himwich and Monica Pate.
\newblock {Infinite Symmetry Algebras in Four-Dimensional Conformal Field
  Theories}, 2026.
\newblock \href {http://arxiv.org/abs/2607.28718} {\path{arXiv:2607.28718}}.

\bibitem{KSZ2022}
Gregory~P. Korchemsky, Emery Sokatchev, and Alexander Zhiboedov.
\newblock {Generalizing event shapes: in search of lost collider time}.
\newblock \href {https://doi.org/10.1007/JHEP08(2022)188} {{\em JHEP}}, 08:188,
  2022.
\newblock \href {http://arxiv.org/abs/2106.14899} {\path{arXiv:2106.14899}}.

\bibitem{KZ2021}
Gregory~P. Korchemsky and Alexander Zhiboedov.
\newblock {On the light-ray algebra in conformal field theories}.
\newblock \href {https://doi.org/10.1007/JHEP02(2022)140} {{\em JHEP}}, 02:140,
  2022.
\newblock \href {http://arxiv.org/abs/2109.13269} {\path{arXiv:2109.13269}}.

\bibitem{KKSZ2020b}
Murat Kologlu, Petr Kravchuk, David Simmons-Duffin, and Alexander Zhiboedov.
\newblock {The light-ray OPE and conformal colliders}.
\newblock \href {https://doi.org/10.1007/JHEP01(2021)128} {{\em JHEP}}, 01:128,
  2021.
\newblock \href {http://arxiv.org/abs/1905.01311} {\path{arXiv:1905.01311}}.

\bibitem{Chicherin2020}
D.~Chicherin, J.~M. Henn, E.~Sokatchev, and K.~Yan.
\newblock {From correlation functions to event shapes in QCD}.
\newblock \href {https://doi.org/10.1007/JHEP02(2021)053} {{\em JHEP}}, 02:053,
  2021.
\newblock \href {http://arxiv.org/abs/2001.10806} {\path{arXiv:2001.10806}}.

\bibitem{Monni2025}
Pier~Francesco Monni, Gherardo Vita, Zhen Xu, and Hua~Xing Zhu.
\newblock {Edge of Safety: Charge-Charge Correlation in the Back-to-Back
  Limit}.
\newblock \href {https://doi.org/10.1103/3pqk-qr64} {{\em Phys. Rev. Lett.}},
  137(3):031901, 2026.
\newblock \href {http://arxiv.org/abs/2508.00977} {\path{arXiv:2508.00977}}.

\bibitem{BSZ2006}
Andrea Banfi, Gavin~P. Salam, and Giulia Zanderighi.
\newblock {Infrared safe definition of jet flavor}.
\newblock \href {https://doi.org/10.1140/epjc/s2006-02552-4} {{\em Eur. Phys.
  J. C}}, 47:113--124, 2006.
\newblock \href {http://arxiv.org/abs/hep-ph/0601139}
  {\path{arXiv:hep-ph/0601139}}.

\bibitem{BK2026}
Carlos Bercini and Sidhaarth Kumar.
\newblock {The simplest event shape in $\mathcal{N}=4$ super Yang--Mills},
  2026.
\newblock \href {http://arxiv.org/abs/2610.18963} {\path{arXiv:2610.18963}}.

\bibitem{CG2000a}
Stefano Catani and Massimiliano Grazzini.
\newblock {Infrared factorization of tree level QCD amplitudes at the
  next-to-next-to-leading order and beyond}.
\newblock \href {https://doi.org/10.1016/S0550-3213(99)00778-6} {{\em Nucl.
  Phys. B}}, 570:287--325, 2000.
\newblock \href {http://arxiv.org/abs/hep-ph/9908523}
  {\path{arXiv:hep-ph/9908523}}.

\bibitem{CaronHuot2018}
Simon Caron-Huot.
\newblock {Resummation of non-global logarithms and the BFKL equation}.
\newblock \href {https://doi.org/10.1007/JHEP03(2018)036} {{\em JHEP}}, 03:036,
  2018.
\newblock \href {http://arxiv.org/abs/1501.03754} {\path{arXiv:1501.03754}}.

\bibitem{BanksZaks1982}
Tom Banks and A.~Zaks.
\newblock {On the Phase Structure of Vector-Like Gauge Theories with Massless
  Fermions}.
\newblock \href {https://doi.org/10.1016/0550-3213(82)90035-9} {{\em Nucl.
  Phys. B}}, 196:189--204, 1982.

\bibitem{CG2000b}
Stefano Catani and Massimiliano Grazzini.
\newblock {The soft gluon current at one loop order}.
\newblock \href {https://doi.org/10.1016/S0550-3213(00)00572-1} {{\em Nucl.
  Phys. B}}, 591:435--454, 2000.
\newblock \href {http://arxiv.org/abs/hep-ph/0007142}
  {\path{arXiv:hep-ph/0007142}}.

\bibitem{Catani1998}
Stefano Catani.
\newblock {The Singular behavior of QCD amplitudes at two loop order}.
\newblock \href {https://doi.org/10.1016/S0370-2693(98)00332-3} {{\em Phys.
  Lett. B}}, 427:161--171, 1998.
\newblock \href {http://arxiv.org/abs/hep-ph/9802439}
  {\path{arXiv:hep-ph/9802439}}.

\bibitem{ChenQQC2026}
Hao Chen, Pier~Francesco Monni, Zhaoyan Pang, Gherardo Vita, and Hua~Xing Zhu.
\newblock {A Conformal Bridge for the Light Transform of QCD Correlation
  Functions}, 2026.
\newblock \href {http://arxiv.org/abs/2603.22388} {\path{arXiv:2603.22388}}.

\bibitem{CaronHuot2023}
Simon Caron-Huot, Murat Kologlu, Petr Kravchuk, David Meltzer, and David
  Simmons-Duffin.
\newblock {Detectors in weakly-coupled field theories}.
\newblock \href {https://doi.org/10.1007/JHEP04(2023)014} {{\em JHEP}}, 04:014,
  2023.
\newblock \href {http://arxiv.org/abs/2209.00008} {\path{arXiv:2209.00008}}.

\bibitem{Low1975}
F.~E. Low.
\newblock {A Model of the Bare Pomeron}.
\newblock \href {https://doi.org/10.1103/PhysRevD.12.163} {{\em Phys. Rev. D}},
  12:163--173, 1975.

\bibitem{Nussinov1975}
S.~Nussinov.
\newblock {Colored Quark Version of Some Hadronic Puzzles}.
\newblock \href {https://doi.org/10.1103/PhysRevLett.34.1286} {{\em Phys. Rev.
  Lett.}}, 34:1286--1289, 1975.

\bibitem{EPSS2001}
Burkhard Eden, Anastasios~C. Petkou, Christian Schubert, and Emery Sokatchev.
\newblock {Partial nonrenormalization of the stress tensor four point function
  in N=4 SYM and AdS / CFT}.
\newblock \href {https://doi.org/10.1016/S0550-3213(01)00151-1} {{\em Nucl.
  Phys. B}}, 607:191--212, 2001.
\newblock \href {http://arxiv.org/abs/hep-th/0009106}
  {\path{arXiv:hep-th/0009106}}.

\bibitem{Eden2012}
Burkhard Eden, Paul Heslop, Gregory~P. Korchemsky, and Emery Sokatchev.
\newblock {Hidden symmetry of four-point correlation functions and amplitudes
  in N=4 SYM}.
\newblock \href {https://doi.org/10.1016/j.nuclphysb.2012.04.007} {{\em Nucl.
  Phys. B}}, 862:193--231, 2012.
\newblock \href {http://arxiv.org/abs/1108.3557} {\path{arXiv:1108.3557}}.

\bibitem{CGP2012}
Miguel~S. Costa, Vasco Goncalves, and Joao Penedones.
\newblock {Conformal Regge theory}.
\newblock \href {https://doi.org/10.1007/JHEP12(2012)091} {{\em JHEP}}, 12:091,
  2012.
\newblock \href {http://arxiv.org/abs/1209.4355} {\path{arXiv:1209.4355}}.

\bibitem{CCP2008}
Lorenzo Cornalba, Miguel~S. Costa, and Joao Penedones.
\newblock {Eikonal Methods in AdS/CFT: BFKL Pomeron at Weak Coupling}.
\newblock \href {https://doi.org/10.1088/1126-6708/2008/06/048} {{\em JHEP}},
  06:048, 2008.
\newblock \href {http://arxiv.org/abs/0801.3002} {\path{arXiv:0801.3002}}.

\bibitem{Besken2021}
Mert Be{\c{s}}ken, Jan de~Boer, and Gr{\'e}goire Mathys.
\newblock {On local and integrated stress-tensor commutators}.
\newblock \href {https://doi.org/10.1007/JHEP07(2021)148} {{\em JHEP}}, 07:148,
  2021.
\newblock \href {http://arxiv.org/abs/2012.15724} {\path{arXiv:2012.15724}}.

\bibitem{CDGP2014}
Miguel~S. Costa, James Drummond, Vasco Goncalves, and Joao Penedones.
\newblock {The role of leading twist operators in the Regge and Lorentzian OPE
  limits}.
\newblock \href {https://doi.org/10.1007/JHEP04(2014)094} {{\em JHEP}}, 04:094,
  2014.
\newblock \href {http://arxiv.org/abs/1311.4886} {\path{arXiv:1311.4886}}.

\bibitem{Lipatov1976}
L.~N. Lipatov.
\newblock {Reggeization of the Vector Meson and the Vacuum Singularity in
  Nonabelian Gauge Theories}.
\newblock {\em Sov. J. Nucl. Phys.}, 23:338--345, 1976.

\bibitem{KLF1977}
E.~A. Kuraev, L.~N. Lipatov, and V.~S. Fadin.
\newblock {The Pomeranchuk Singularity in Nonabelian Gauge Theories}.
\newblock {\em Sov. Phys. JETP}, 45:199--204, 1977.

\bibitem{BL1978}
I.~I. Balitsky and L.~N. Lipatov.
\newblock {The Pomeranchuk Singularity in Quantum Chromodynamics}.
\newblock {\em Sov. J. Nucl. Phys.}, 28:822--829, 1978.

\bibitem{DDEHPS2013}
James Drummond, Claude Duhr, Burkhard Eden, Paul Heslop, Jeffrey Pennington,
  and Vladimir~A. Smirnov.
\newblock {Leading singularities and off-shell conformal integrals}.
\newblock \href {https://doi.org/10.1007/JHEP08(2013)133} {{\em JHEP}}, 08:133,
  2013.
\newblock \href {http://arxiv.org/abs/1303.6909} {\path{arXiv:1303.6909}}.

\bibitem{RV2000}
E.~Remiddi and J.~A.~M. Vermaseren.
\newblock {Harmonic polylogarithms}.
\newblock \href {https://doi.org/10.1142/S0217751X00000367} {{\em Int. J. Mod.
  Phys. A}}, 15:725--754, 2000.
\newblock \href {http://arxiv.org/abs/hep-ph/9905237}
  {\path{arXiv:hep-ph/9905237}}.

\bibitem{BPST2007}
Richard~C. Brower, Joseph Polchinski, Matthew~J. Strassler, and Chung-I Tan.
\newblock {The Pomeron and gauge/string duality}.
\newblock \href {https://doi.org/10.1088/1126-6708/2007/12/005} {{\em JHEP}},
  12:005, 2007.
\newblock \href {http://arxiv.org/abs/hep-th/0603115}
  {\path{arXiv:hep-th/0603115}}.

\bibitem{GLS2015}
Nikolay Gromov, Fedor Levkovich-Maslyuk, and Grigory Sizov.
\newblock {Quantum Spectral Curve and the Numerical Solution of the Spectral
  Problem in AdS5/CFT4}.
\newblock \href {https://doi.org/10.1007/JHEP06(2016)036} {{\em JHEP}}, 06:036,
  2016.
\newblock \href {http://arxiv.org/abs/1504.06640} {\path{arXiv:1504.06640}}.

\bibitem{HuPasterski2023}
Yangrui Hu and Sabrina Pasterski.
\newblock {Celestial conformal colliders}.
\newblock \href {https://doi.org/10.1007/JHEP02(2023)243} {{\em JHEP}}, 02:243,
  2023.
\newblock \href {http://arxiv.org/abs/2211.14287} {\path{arXiv:2211.14287}}.

\bibitem{HMOPS2026}
Enrico Herrmann, Ian Moult, Brett Oertel, Julio Parra-Martinez, and Biswajit
  Sahoo.
\newblock {Gravitational Memory Correlators and BMS Symmetry}, 2026.
\newblock \href {http://arxiv.org/abs/2611.18827} {\path{arXiv:2611.18827}}.

\bibitem{Belin2021}
Alexandre Belin, Diego~M. Hofman, Gr{\'e}goire Mathys, and Matthew~T. Walters.
\newblock {On the stress tensor light-ray operator algebra}.
\newblock \href {https://doi.org/10.1007/JHEP05(2021)033} {{\em JHEP}}, 05:033,
  2021.
\newblock \href {http://arxiv.org/abs/2011.13862} {\path{arXiv:2011.13862}}.

\bibitem{Allendes2013}
Pedro Allendes, Bernd Kniehl, Igor Kondrashuk, Eduardo~A. {Notte Cuello}, and
  Marko {Rojas Medar}.
\newblock {Solution to Bethe-Salpeter equation via Mellin-Barnes transform}.
\newblock \href {https://doi.org/10.1016/j.nuclphysb.2013.01.012} {{\em Nucl.
  Phys. B}}, 870:243--277, 2013.
\newblock \href {http://arxiv.org/abs/1205.6257} {\path{arXiv:1205.6257}}.

\bibitem{BhatMaity2023}
Faizan Bhat and Pronobesh Maity.
\newblock {Perturbative Mellin amplitudes of $\mathcal{N}=4$ SYM}, 2023.
\newblock \href {http://arxiv.org/abs/2312.17692} {\path{arXiv:2312.17692}}.

\end{thebibliography}

\end{document}